\documentclass[12pt,preprint]{aastex}
\usepackage{graphicx-psmin}
\begin{document}


\newcommand{\kms}        {km s$^{-1}$}
\newcommand{\herschel} {\textit{Herschel}}
\newcommand{\water} {H$_2$O}
\newcommand{\hcop} {HCO$^+$}
\newcommand{\den}{${n}_{\mathrm{H}_2}$}
\newcommand{\ncolone}{$N_{\mathrm{CO}}$}
\newcommand{\ncoltwo}{$N(^{13}$CO)}
\newcommand{\tkin}{$T_{\mathrm{kin}}$}
\newcommand{\cmtwo}{cm$^{-2}$}
\newcommand{\cmthree}{cm$^{-3}$}
\newcommand{\coone}{$^{12}$CO}
\newcommand{\cotwo}{$^{13}$CO}
\newcommand{\msun}{M$_{\odot}$}
\newcommand{\lsun}{L$_{\odot}$}
\newcommand{\mic}{$\mu$m}
\newcommand{\htwo}{H$_{2}$}
\newcommand{\lfir} {$L_{\mathrm{FIR}}$}

\title{Observations of Arp~220 using \textit{Herschel}-SPIRE: An Unprecedented View of the Molecular Gas in an Extreme Star Formation Environment}
\author{Naseem Rangwala\altaffilmark{1}, Philip R. Maloney\altaffilmark{1}, Jason Glenn\altaffilmark{1}, Christine D. Wilson\altaffilmark{4}, Adam Rykala\altaffilmark{2}, Kate Isaak\altaffilmark{3}, Maarten Baes\altaffilmark{13}, George J. Bendo\altaffilmark{8}, Alessandro Boselli\altaffilmark{14}, Charles M. Bradford\altaffilmark{5}, D. L. Clements\altaffilmark{6}, Asantha Cooray\altaffilmark{12}, Trevor Fulton\altaffilmark{7}, Peter Imhof\altaffilmark{7}, Julia Kamenetzky\altaffilmark{1}, Suzanne C. Madden\altaffilmark{11}, Erin Mentuch\altaffilmark{4}, Nicola Sacchi\altaffilmark{9}, Marc Sauvage\altaffilmark{11}, Maximilien R. P. Schirm\altaffilmark{4}, M. W. L. Smith\altaffilmark{2},  Luigi Spinoglio\altaffilmark{9}, Mark Wolfire\altaffilmark{10}}

\altaffiltext{1}{Center for Astrophysics and Space Astronomy, University of Colorado, 1255 38th street, Boulder, CO 80303} 
\altaffiltext{2}{School of Physics \& Astronomy, Cardiff University, Queens Buildings The Parade, Cardiff CF24 3AA, UK}
\altaffiltext{3}{ESA Astrophysics Missions Division, ESTEC, PO Box 299, 2200 AG Noordwijk, The Netherlands}
\altaffiltext{4}{Dept. of Physics \& Astronomy, McMaster University, Hamilton, Ontario, L8S 4M1, Canada }
\altaffiltext{5}{JPL, Pasadena, CA 91109}
\altaffiltext{6}{Astrophysics Group, Imperial College, Blackett Laboratory, Prince Consort Road, London SW7 2AZ, UK}
\altaffiltext{7}{Blue Sky Spectroscopy Inc, Suite 9-740 4th Avenue South, Lethbridge, Alberta T1J 0N9, Canada}
\altaffiltext{8}{UK ALMA Regional Centre Node, Jordell Bank Center for Astrophysics, School of Physics and Astronomy, University of Manchester, Oxford Road, Manchester M13 9PL, U.K.}
\altaffiltext{9}{Istituto di Fisica dello Spazio Interplanetario, INAF, Via del Fosso del Cavaliere 100, I-00133 Roma, Italy}
\altaffiltext{10}{Dept. of Astronomy, University of Maryland, College Park, MD 20742}
\altaffiltext{11}{CEA, Laboratoire AIM, Irfu/SAp, Orme des Merisiers, F-91191 Gif-sur-Yvette, France}
\altaffiltext{12}{Department of Physics \& Astronomy, University of California, Irvine, CA 92697, USA}
\altaffiltext{13}{Sterrenkundig Observatorium, Universiteit Gent, Krijgslaan 281 S9, B-9000 Gent, Belgium}
\altaffiltext{14}{Laboratoire d'Astrophysique de Marseille, UMR6110 CNRS, 38 rue F. Joliot-Curie, F-13388 Marseille France}

\begin{abstract}
We present Herschel SPIRE-FTS observations of Arp~220, a nearby ultra-luminous infrared galaxy. The FTS provides continuous spectral coverage from 190 -- 670 \mic, a wavelength region that is either very difficult to observe or completely inaccessible from the ground. The spectrum provides a good measurement of the continuum and detection of several molecular and atomic species. We detect luminous CO (J = 4-3 to 13-12) and water rotational transitions with comparable total luminosity $\sim 2 \times 10^{8}$ $L_\odot$; very high-J transitions of HCN (J = 12-11 to 17-16) in absorption; strong absorption features of rare species such as OH$^{+}$, \water$^{+}$, and HF; atomic lines of [\ion{C}{1}] and [\ion{N}{2}].

The modeling of the continuum shows that the dust is warm, with $T = 66$~K, and has an unusually large optical depth, with $\tau_\mathrm{dust} \sim 5$ at 100~\mic. The total far-infrared luminosity of Arp~220 is $L_\mathrm{FIR} \sim 2 \times 10^{12}$~L$_\odot$.

Non-LTE modeling of the extinction corrected CO rotational transitions shows that the spectral line energy distribution of CO is fit well by two temperature components: cold molecular gas at $T \sim 50$~K and warm molecular gas at $T \sim 1350^{+280}_{-100}$ ~K (the inferred temperatures are much lower if CO line fluxes are not corrected for dust extinction). These two components are not in pressure equilibrium. The mass of the warm gas is 10\% of the cold gas, but it dominates the CO luminosity. The ratio of total CO luminosity to the total FIR luminosity is $L_\mathrm{CO}/L_\mathrm{FIR} \sim 10^{-4}$ (the most luminous lines, such as J = 6-5, have $L_\mathrm{CO, J = 6-5}/L_\mathrm{FIR} \sim 10^{-5}$). The temperature of the warm gas is in excellent agreement with the observations of \htwo\ rotational lines. At 1350~K, \htwo\ dominates the cooling ($\sim 20$~$L_\odot/M_\odot$) in the ISM compared to CO ($\sim 0.4$~$L_\odot/M_\odot$). We have ruled out PDR, XDR and cosmic rays as likely sources of excitation of this warm molecular gas, and found that only a non-ionizing source can heat this gas; the mechanical energy from supernovae and stellar winds is able to satisfy the large energy budget of $\sim 20$ $L_\odot/M_\odot$. Analysis of the very high-J lines of HCN strongly indicates that they are solely populated by infrared pumping of photons at 14\mic. This mechanism requires an intense radiation field with $T > 350$~K. 

We detect a massive molecular outflow in Arp~220 from the analysis of strong P Cygni line profiles observed in OH$^{+}$, \water$^{+}$, and \water. The outflow has a mass $\gtrsim 10^{7}$ $M_\odot$ and is bound to the nuclei with velocity $\lesssim 250$ \kms. The large column densities observed for these molecular ions strongly favor the existence of an X-ray luminous AGN ($10^{44}$ ergs s$^{-1}$) in Arp 220. 
\end{abstract}


\section{Introduction}
 Molecular gas is the raw material for the formation of stars within galaxies. However, the star formation process inputs energy and momentum back into this gas - and into the ISM in general - through radiation, stellar winds, and supernovae. The impact of the feedback from both star formation and AGN activity on the ISM can be profound \citep{maloney99}, and the resulting change in the physical state of the gas must in turn affect the star formation process. In the general framework of galaxy evolution, disentangling the relative contributions of these processes for high-redshift objects, however, is an extremely difficult undertaking, due to their small angular size and faintness; in addition, many luminous 
high-$z$ sources are heavily obscured by dust, ruling out most of the traditional approaches for studies of galaxy energetics. Nearby galaxies such as M82, Arp~220 and NGC~1068, for which deep and relatively high resolution observations can be obtained at multiple wavelengths, are commonly used templates for calibrating the
observations of high-\textit{z} luminous dusty galaxies. 

The study of the star formation process and of the effect of AGN activity in nearby galaxies can be done in the far-infrared (FIR) and submillimeter (submm) spectral range, where a large variety of high-J rotational molecular transitions occur (CO, HCN, HCO$^+$ etc.). These transitions have a large span in critical densities making them excellent tracers of the physical conditions of gas over a wide range in temperatures (10 -- 1000 K) and densities ($10^3$ -- $10^8$ \cmthree). Also, their line ratios can be used as diagnostics for distinguishing between different energy sources responsible for the excitation of the gas, e.g., starbursts or AGN. The low-J transitions (up to J = 4-3) of these molecules trace the cooler gas and have been observed from ground-based radio and submm telescopes. These low-J transitions cannot distinguish between the AGN and starburst systems. However, the mid-J to higher-J transitions, which can trace warmer and denser gas and be used to test models that distinguish between AGN and starbursts \citep[e.g.][]{spaans08}, are either difficult to observe or completely inaccessible from the ground. Until the advent of \herschel\ \citep{pilbratt10} they had only been observed in a handful of extragalactic objects.

The Very Nearby Galaxy Survey (VNGS) (PI: C. Wilson) is one of the guaranteed-time key projects of \herschel\ that is using both the Spectral and Photometric Imaging Receiver (SPIRE) \citep{griffin2010}, and the Photodetector Array Camera and Spectrometer (PACS) \citep{poglitsch10} instruments. This paper focuses only on the data from the SPIRE instrument, which is being used to obtain high S/N photometric and spectroscopic maps of a sample of 13 nearby archetypical galaxies. The imaging spectrometer on SPIRE is a Fourier Transform Spectrometer (FTS) \citep{naylor2010}, which provides continuous spectral coverage from 190 -- 670 $\mu$m ($\sim$ 450 -- 1550 GHz). This region covers the spectral features of several key molecular and atomic species that are powerful diagnostic of the physics and chemistry of the ISM. All the galaxies in this survey will have complementary PACS observations either as a part of this program or another key project, allowing exploration of a full spectral view from $50 - 700 \mu$m. In this paper, we present the SPIRE observations of Arp~220. Until now only two nearby galaxies observed with SPIRE-FTS have been published. They are M82 \citep{panuzzo2010} and Mrk~231 \citep{pwerf2010}. In both cases the mid-J to high-J CO ladder was detected. Modeling of this CO ladder showed that in M82 the excitation of molecular gas was governed by the starburst, whereas in Mrk~231 the excitation was dominated by the central luminous AGN.

Arp~220 is the nearest Ultra Luminous Infrared galaxy (ULIRG) at a distance of about 77 Mpc and $z \sim 0.0181$. It has $L_{\mathrm{FIR}} \sim 10^{12}$ L$_{\odot}$, and is one of the most popular templates for studies of high-z dusty  galaxies. It has  two merging nuclei separated by $\sim 1^{\prime\prime}$ ($\sim 360$ pc at $D_{\mathrm{Arp220}} = 77$ Mpc) \citep[][SC97 hereafter]{scoville97} and together they have a large reservoir of molecular gas ($\sim 10^{10}$ M$_{\odot}$). Arp~220 has been observed extensively over the years across the electromagnetic spectrum. The extreme star formation environment in this galaxy provides an excellent laboratory to understand the processes affecting star formation and possibly AGN feedback. 

In this paper, we describe the state of the molecular gas by analyzing and modeling the new detections of high-J CO and HCN lines. We discuss sources of energy that could be responsible for the excitation of various molecular species detected in this spectrum. The issue of a hidden AGN has been long debated for Arp~220. We address this issue using the detection of rare molecular species that can only be detected by \herschel. We also present results from dust modeling by using the continuum measurements from SPIRE-FTS and SPIRE-photometer that adds data at wavelengths not accessible before. We summarize these results and discuss (in light of the new results) Arp~220 as template for broadband spectral energy distributions for high-z submm galaxies.
\section{Observations and Data Reduction}
Arp 220 was observed with the SPIRE-FTS in a high spectral resolution with FWHM $\sim$ 1.44 GHz (or ranging from 950 \kms\ to 280 \kms\ across the spectral range), single pointing and sparse image sampling mode  on 13 February, 2010. The total on-source integration time was 10,445~seconds. We used a deep dark sky observation with total integration time of 13,320~seconds taken on 21 November, 2010 to account for the emission from the telescope. The FTS has two detectors arrays, called SLW and SSW, to cover the long and short wavelength ranges respectively, with a small overlap in wavelength. Arp~220 was also observed with the SPIRE photometer on 28 December, 2009 with a total on-source integration time of 214s at 250 \mic, 350\mic\ and 500 \mic. The calibration uncertainties in SPIRE-FTS are about 10\% at the lower end of the frequency range but are as good as 5\% across the rest of the spectrum. For the SPIRE photometer, the systematic uncertainty, which is related to the model for Neptune, is 5\%.  This is correlated among the three SPIRE bands. The random uncertainty, which is related to the ability of the instrument to repeat measurements of Neptune, is $\sim 2\%$. This is independent for each of the SPIRE bands.
 
We processed the SPIRE FTS data using the Herschel data processing pipeline \citep{fulton2010}, HIPE version 6.0. The spectrum is shown in Figure \ref{spec} and displays both emission and absorption features. Arp~220 is a point source for the diffraction limited SPIRE beam size\footnote{The SPIRE beam shapes are not gaussian; the effective beam solid angle can be found in the \herschel\ Observer's manual.}, which ranges between FWHM of $17^{\prime\prime} - 40^{\prime\prime}$. The SLW and SSW continuum match perfectly across the overlap region. The flux densities from the SPIRE photometer were measured by fitting a two dimensional elliptical Gaussian function to the source in the timeline data.  In this fit, the background was treated as a free parameter in the fit.  The fit was performed to a inner region with radii of 22\arcsec, 30\arcsec, and 42\arcsec\ in the 250, 350, and 500 \mic\ bands, respectively (which were selected because they contain the part of the beam profile that appears Gaussian and exclude the Airy rings) and using a background annulus with radii between 300\arcsec\ and 350\arcsec. The peak of the best fitting Gaussian function in each wave band corresponds to the flux density at that wavelength. The three SPIRE photometer measurements are overplotted in Figure \ref{spec} as red solid circles. They are in very good agreement with the FTS continuum within the statistical and calibration uncertainties of SPIRE-FTS and SPIRE-photometer.

The spectral line shape of the FTS can be modeled to a good accuracy by a sinc function with a FWHM of about 1.44 GHz. The ground-based observations measure large line velocity widths, of the order of 500 \kms\ in CO and HCN  in Arp~220 (SC97) \citep[also see][]{greve09}. Therefore,  in our FTS spectrum we start resolving the broad spectral lines towards the higher frequency end. We use a sinc function of variable FWHM to fit the emission and absorption lines using the MPFIT package in IDL. In Table 1 we report the integrated line fluxes in Jy \kms\ for emission lines and equivalent widths in $\mu$m for the absorption lines along with their 1$-\sigma$ statistical uncertainties. The observed spectral line shape suffers from small asymmetries from instrumental effects, and the line shapes could be affected by blending from weak lines. The effect on the line fluxes from asymmetries is expected to be $\lesssim  2\%$. 

\section{Line Identifications}\label{lineid}
The spectrum (Fig.\,1)  shows the FIR continuum and the detection of several key molecular and atomic species and Figure \ref{specmulti} shows zoom-in views of some of the spectral lines discussed in this section. In Table 1, we list their transitions, rest frequencies and integrated fluxes. The FWHMs mentioned in this section are from our sinc line fitting. 

We detect a luminous CO  emission ladder from J = 4-3 to J = 13-12 and several water transitions with total water luminosity comparable to CO. Compared to the velocity widths measured in CO from the ground-based observations \citep[e.g. SC97,][]{greve09}, the higher-J CO lines in our spectrum are much narrower. These high-J CO lines are not resolved and follows the resolution of FTS, which at the frequency of J = 12-11 line is $\sim 310$ \kms. This suggests that either the high-J CO emission is coming from only one of the nuclei or that it is a dynamically separate component from the low-J lines. On the other hand, the water lines are much broader with velocity FWHM of $\sim 500$ \kms\, implying that this emission could be coming from both nuclei. The observed velocity separation between the two nuclei in Arp~220 varies considerably from 50 \kms\ (Downes \& Solomon 1998; Sakamoto et al 2008) to 250 \kms (Scoville et al. 1997; Sakamoto et al 1999). Therefore, it is difficult to conclude whether the line widths are affected by the velocity shift between the two nuclei or by the dynamics of the molecular material.

In addition to the emission line features, the spectrum also shows several absorption features. Most surprising are the detections of  five very high-J HCN (J = 12-11 to J = 17-16) lines in absorption; their low-J transitions (J = 1-0 to J = 4-3) are detected in emission with ground-based observatories \citep [and references therein]{greve09}. Such high-J transitions of HCN have never been detected before in external galaxies, in emission or absorption.

Strong absorption lines are detected from hydrides, including three transitions of OH$^{+}$ (see Figure \ref{pcygni}) and one of  CH$^{+}$ (see Figure \ref{specmulti} (e)). We report four detections of \water$^{+}$ transitions; two of them (1115, 1139 GHz) show strong absorption and the other two, at 742 and 746 GHz \citep[rest frequencies are reported in][]{bruderer06, ossenkopf10}, appear in emission. Before \herschel, \water$^{+}$ was only detected in comets. The transitions at 1115 and 1139 GHz (see Figure \ref{pcygni}) were first detected in Galactic molecular clouds and in extragalactic objects by HIFI \citep{ossenkopf10} and SPIRE-FTS \citep{pwerf2010}. We have the first detection of transitions at 742 and 746 GHz (see Figure \ref{specmulti}(c)). The detections of OH$^{+}$ and \water$^{+}$ are very important in Arp~220 for modeling the water transitions because they are the major intermediaries in the ion-neutral chemistry network producing water in the ISM \citep{gerin2010,neufeld2010}, and their detection is very difficult from current ground-based observatories. \herschel\ HIFI and SPIRE-FTS also made the first detection of the HF 1232 GHz line \citep{neufeld2010II, pwerf2010}, and we detect it in absorption in Arp~220. The spectrum also shows several features of Nitrogen hydrides including NH, NH$_2$ and NH$_3$.

The rest frequencies of all the above detections of the molecular species are consistent with the systemic $z = 0.0181$ of Arp~220. However we detect two emission features at $\sim 532.6$ and $\sim 536.8$ GHz (see Figure \ref{specmulti} (a)) that are close in frequency to the J = 6-5 transitions of HCN and HCO$^{+}$; their low-J transitions (up to J = 4-3) have been seen in emission from the ground. These features also have larger FWHM ($\sim 1000$ \kms) compared to the spectral resolution at these frequencies.
The redshifted frequencies of both HCN and HCO$^{+}$ are offset from the systemic velocity of Arp~220 by $\sim -0.9$ GHz (-500 \kms) and $\sim -1.7$ GHz (-1000 \kms), respectively. Such large shifts are not seen for any other detections in our spectrum. Since we see very high-J lines of HCN in absorption it is possible that the HCN J = 6-5 line is partially absorbed by the same gas producing HCN in absorption at higher frequencies. This effect could shift its apparent central frequency. The case for HCO$^{+}$ is even more problematic. The rest frequency of the emission feature is in perfect agreement with the rest frequency of HOC$^{+}$ molecule instead of HCO$^{+}$.  Since the width of this feature is about $\sim 1000$ \kms,  these two molecules could be blended. However, identifying this feature as HOC$^{+}$ would imply that it has comparable to or higher line strength than HCO$^{+}$. This would be extremely unusual and has never been observed in other galaxies or in Galactic molecular clouds. All observations of HOC$^{+}$ and HCO$^{+}$ show that the latter is at least 100 - 1000  times stronger \citep{martin09, savage04}. In Table 1 we mark these identifications with a question mark.

Among the atomic species we detect [\ion{C}{1}] and [{\ion{N}{2}] lines (see Figure \ref{specmulti} (d) \& (i)). The detection of the two [\ion{C}{1}] lines at 492 and 809 GHz have been rare in the past. The forbidden [\ion{N}{2}] line at 1462 GHz has been inaccessible from ground and rarely accessible from space. The \herschel\ FTS has been very successful in detecting them in several Galactic and extragalactic targets. The [\ion{C}{1}] (1-0) 492 GHz line survey for nearby galaxies was done by Gerin \& Phillips (2000), in which this line was detected in Arp~220.  We report the first detections of [\ion{C}{1}] (2-1) and [\ion{N}{2}] lines in Arp~220.  
The [\ion{N}{2}] line width is broader ($\sim 400$ \kms) than expected from the FTS resolution of about 280 \kms\ at these high frequencies, implying that we might be resolving the [\ion{N}{2}] emission from the two Arp~220 nuclei. More discussion of these two atomic lines is presented in Section \ref{atomic}.
 

\section{Dust in Arp~220}\label{dustext}
In addition to line fluxes, the FTS also provides a good measurement of the continuum at wavelengths that have not been accessible before.  We model the dust spectral energy distribution (SED) by using the combination of FTS continuum data (listed in Table 2) and other data in the literature, providing wide wavelength coverage. The modeling yields constraints on dust temperature ($T$), spectral index ($\beta$) and the frequency ($\nu_{0}$) at which the dust optical depth becomes unity. 

We combined FTS continuum with ISO-LWS continuum data \citep{brauher08} providing a wavelength coverage from $50 - 670~\mu$m. We chose ISO-LWS over other observations because it measures several data points covering the peak of the SED. In addition, the calibration of the two data sets agree extremely well. We also added three continuum data points from the SPIRE photometer at 250~\mic, 350~\mic\ and 500~\mic.  We fit these data with the following modified blackbody model:
\begin{equation}
f_{\nu} \propto B_{\nu} (1 - \mathrm{e}^{-\tau}), 
\end{equation} 
where\\
\begin{equation}
\tau ={\large (\frac{\nu}{\nu_{0}})^{\beta}}  \,\,\, \& \quad   B_{\nu}\,(T) = \frac{2h\nu^{3}}{c^2}\frac{1}{\mathrm{e}^{\frac{h\nu}{kT}} - 1}.
\end{equation}

 We generated a grid of models over a range in $T$, $\beta$ and $\nu_{0}$. Using this grid we produced likelihood distributions of $T$, $\beta $ and $\nu_0$ by comparing the model continuum fluxes to the observed ones (corrected for $z_\mathrm{Arp220} = 0.0181$). In addition to statistical uncertainties we also included calibration uncertainties using a covariance matrix, a much better approach than adding them in quadrature. Figure \ref{dustsed} shows the highest likelihood model fit to the SPIRE-FTS (blue - open circles), SPIRE-photometer (red - open triangles; listed in Table 2) and ISO-LWS (red - open squares) data. The reduced chi-squared of this fit is 0.5, indicating that our calibration errors are likely too conservative. This excellent fit to the data is consistent with only a single temperature component. We overplot data from IRAS in green-solid squares \citep{sanders03}, SCUBA in green - solid triangles \citep{dunne00}, ISOPHOT in purple - open diamonds \citep{klaas01} and Planck in orange open triangles (Planck Collaboration 2011, arXiv:1101.2041), which also show good agreement with the fit down to 15 $\mu$m, below which the single temperature component fit breaks down. The contours in Figure \ref{dust} show 68\%, 95\% and 99\% confidence levels for the three model parameters.  The addition of SPIRE data provides far better constraints on the dust parameters, as indicated by the compact contours, than was possible with previous measurement alone. The highest likelihood values and their 1--$\sigma$ ranges are: $T$ = 66.7~K (66.4 -- 66.9), $\beta = 1.83$ (1.82 -- 1.85), and $\nu_0 = 1277 $ GHz (1262 -- 1287). Generally, the value of the emissivity index, $\beta$, is fixed at 2.0 for the silicate and graphite dust \citep{draine84}. Treating $\beta$ as a free parameter in our analysis gives a value of $\sim 1.8$, which is very close to the model value. The effect of dust extinction in Arp~220 is very large. The dust optical depth is unity at $\sim 240$~\mic\ and $\sim 5.0$ at 100~$\mu$m. The total hydrogen column density is $\sim 10^{25}$ \cmtwo\ using the 250 $\mu$m dust cross-section from \citet{li01}, which has a systematic uncertainty close to a factor of 3. All the emission line fluxes in our spectrum need to be corrected for this large dust extinction. We estimate a dust mass, $M_\mathrm{dust} \sim 10^{8}\,M_\odot$, for this model using the flux at 250 \mic\ and $\kappa_{250} = 0.517$ m$^{2}$ kg$^{-1}$ \citep{li01}. This gives the gas-to-dust mass ratio\footnote{The molecular to atomic gas mass fraction is much higher for ULIRGs \citep{mirabel89} compared to quiescent galaxies, so we ignore the contribution of the atomic gas while calculating the gas-to-dust mass ratio.} of $\sim 100$ using the molecular gas mass of $\sim 10^{10}\,M_\odot$ for Arp~220 measured by \citet{scoville97}.
 
 The large dust optical depths at submm and FIR wavelengths in Arp~220 are not unexpected. There is a high column density of molecular gas measured from CO line observations (e.g. SC97). The FIR spectrum shows both [OI] 63 \mic\ and [\ion{N}{2}] 122 \mic\ lines in absorption \citep{alfonso04, fischer00}, compared to other galaxies where these lines usually appear in emission, suggesting a stronger dust continuum in Arp~220. The [\ion{C}{2}] 158 \mic\ line deficit relative to the FIR continuum ($L_{\mathrm{C}^{+}}$/$L_{\mathrm{FIR}} \sim 2 \times 10^{-4}$) in this galaxy is one of the most extreme compared to other ULIRGs \citep[e.g.,][] {luhman03}. From an interferometric map of Arp~220 at 860 \mic, \citet{sakamoto08} estimated a large dust optical depth, of $\sim 1$, even at this wavelength. In addition, if the observed CO ladder is not corrected for the large dust extinction (see section \ref{comodel}), then the temperature derived for the molecular gas traced by CO from our radiative transfer modeling will be in large disagreement with the temperature derived from the mid-IR \htwo\ lines observed from \textit{Spitzer}. Several previous works have argued for large dust optical depths in this galaxy \citep[e.g.,][]{lisenfeld00, fischer00, klaas01, alfonso04, downes07, papad10}. Our result of large dust optical depth is consistent with all these studies. However, the addition of continuum data from SPIRE has reduced the uncertainties on the dust parameters by an order of magnitude compared to the previous studies. 
 
Multiple studies have shown that dust SEDs for ULIRGs may not only be fit by a single temperature optically-thick dust component but also by multiple optically-thin (assuming $\tau_\mathrm{dust} << 1$) temperature components, with both models fitting the SED equally well \citep{klaas01, blain03}. However we found that fitting an optically-thin two temperature component model to the Arp~220 continuum data did not produce realistic results. Our optically-thin 2-component model fit gives $\sim 25$ K and $\sim 46$ K for the cold and warm temperature components, and beta = 2. Even though these temperatures are realistic the dust mass for this model is not: it yields an estimated dust mass of $\sim 10^{9}$ M$_\odot$, using the flux at 250 \mic. A \citet{draine07} model fit for an optically thin dust assumption also produces a large dust mass ($\sim 6 \times 10^{8}$ M$_\odot$). This is an order of magnitude larger than the optically thick result of $\sim 10^{8}$ M$_\odot$. For M(H$_{2}) \sim 10^{10}$ M$_\odot$ in Arp~220 (SC97), the gas-to-dust mass ratio is only 10 in the optically thin case. Using our preferred value of M(H$_{2}) = 5.2 \times 10^{9}$ M$_\odot$ derived from modeling of the CO line energy distribution (see section 5 and table 3), the gas-to-dust mass ratio is still only $\sim 5$ in the optically thin case; using larger value estimated by SC97 would only raise it by a factor of 2. Even after considering a factor of three uncertainty in $\kappa_{250}$, this ratio is extremely low \citep[expected to be $\gtrsim 100$, e.g., ][]{hildebrand83, devereux90, sanders91}. Therefore, modeling the dust with a general modified blackbody model in which no prior assumption is made about dust optical depth is strongly favored.

Historically, fitting a single thermal component to model dust emission between 24 \mic\ and 600 \mic\ has been considered unrealistic. This is because the dust is expected to be distributed at different distances from the heating source; the dust observed at 60 \mic\ would be much closer to the heating source than the dust observed at 500 \mic.  Also, even for dust heated by a uniform radiation field, transiently-heated smaller dust grains will typically produce emission that peaks at shorter wavelengths, while the larger dust grains will appear to be in thermal equilibrium at ~15-30 K \citep[e.g.,][]{li01, draine07}. However, in case of Arp~220, the dust is optically thick at these shorter wavelengths, and so transiently-heated small grains as well as grains closer to the heating source are not visible. From the average dust measurements we find that the FIR luminosity is dominated by a skin at $\sim 66$~K. 
 
We estimate a total FIR luminosity $L_{\mathrm{FIR}_{(8 - 1000\mu \mathrm{m}})} = 1.77 \times 10^{12}$ $L_\odot$ by integrating the model SED.  Using this value of  L$_{\mathrm{FIR}}$ and $T = 66$ K in the Stefan-Boltzmann relation we obtain a source size of about 240 pc radius ($\sim 0.64^{\prime\prime}$ using $D_{\mathrm{Arp220}} = 77$ Mpc). This gives a source solid angle of $4.1 \times 10^{-11}\,\mathrm{sr}$, which is a factor of $\sim 1.6$ smaller than measured for the CO (1-0) line by SC97. Our single temperature fit suggests that the dust emission is coming from one nucleus alone. This is also supported by our high-J CO line observations, which have much smaller line velocity widths compared to the low-J CO lines, implying that the high-J CO emission is predominantly coming from one nucleus. This conclusion is inconsistent with the submm interferometric observations from \citep{sakamoto08} (860 \mic) and \citep{matsushita09} (435 \mic). These studies suggest that the dust emission in the western nucleus is more compact than the eastern nucleus and the luminosity of the western nucleus is either comparable to or higher (2-3 times) than the eastern nucleus. The sizes and the emission of the two nuclei are likely to change with wavelength such that this trend may not hold over the wavelength range that we are observing.  
\section{CO: Spectral Line Energy Distribution and Non-LTE modeling}\label{comodel}
 The CO line luminosities are affected by the large dust optical depths derived in the previous section. We correct for this dust extinction by assuming that the dust and gas are well mixed; i.e., $I  = I_{0} (1 - \mathrm{e}^{-\tau_{\lambda}})/\tau_{\lambda}$ \citep[][and references therein]{haas01}. Figure \ref{cosed} (left panel) shows the observed extinction corrected luminosity distribution of the  CO ladder from J = 1-0 to J = 13-12. The extinction correction factor ($I_{0}/I$) ranges from 1.08 at 450 GHz to 1.95 at 1600 GHz. The solid circles are the FTS measurements from this work and the blue triangles are the average of the ground-based observations compiled by \citet{greve09}. In addition, we show new observations from the JCMT \citep{papad10} of CO (6-5) line and from Z-Spec (private comm.) of CO (2-1) line for comparison. The JCMT flux is lower by a factor of 1.3, likely missing some flux owing to large velocity widths in Arp 220.  The Z-spec CO (2-1) measurement is higher than all the other ground-based measurements listed in \citet{greve09} but in agreement with the JCMT CO (2-1) measurement within their 1$-\sigma$ uncertainties. \citet{matsushita09} obtained an interferometric map of CO J = 6-5 line using SMA. Figure 7 of their paper indicates that the CO J = 6-5 emission is mainly coming from only one of the nuclei compared to the low-J CO transitions (J = 1-0 to J = 3-2), which clearly show emission from both nuclei. This is consistent with our observations of the narrower line widths of the high-J CO transitions.
However, \citet{matsushita09} are missing $\sim 70$\% of the flux in this line, obtaining integrated line flux of only $1250 \pm 250$ Jy \kms.

The total CO luminosity is about $2 \times 10^{8}\, L_{\odot}$ and is dominated by the mid-J to high-J CO transitions.  The shape of the CO spectral line energy distribution (SLED) in Arp~220 is similar to that of M82 (P10), a starburst galaxy,  in which the CO line fluxes are highest for the mid-J lines and then fall off at higher-J. This is in contrast to Mrk~231 (Van der Werf et al. 2010), in which the CO SLED rises up to J = 5-4 and remains flat for the higher-J transitions, which can be explained by the presence of a central X-ray source illuminating the circumnuclear region. The shape of the CO SLED has been suggested as a discriminant between star formation dominated versus XDR dominated systems \citep{werf09}. It will be possible to test this in the near future by compiling CO SLEDs for a sample of nearby galaxies observed by \herschel\ from several key projects. However, a practical application of using the shape of the CO SLED as a discriminant may be complicated in systems where the luminosities of star formation and AGN are comparable. In Arp~220, the shape of the CO SLED is consistent with star formation but we show later (in Section 8) that there is also evidence for a luminous AGN from observations of other molecules in the FTS spectrum. This is discussed further in Section 9.
\subsection{Non-LTE Modeling}
We used the full observed CO SLED from J = 1-0 to J = 13-12, with the exception of the J = 10-9 line, which is blended with a water line (see Figure \ref{specmulti} (f)),  to model the CO excitation and radiative transfer using a variant of the non-LTE code, RADEX \citep{tak07}. This code computes the intensities of molecular lines by iteratively solving for
statistical equilibrium, using an escape probability formalism. In our case we used the escape probability formalism for an expanding spherical shell (large velocity gradient, LVG\footnote{The form of escape probability for LVG and uniform sphere models are very similar and hence either choice will not introduce any significant difference in the CO flux values.}). 
The code starts from an optically thin case to generate the initial guess for the level populations, then iterates until a self consistent solution is achieved such that the optical depths of the lines are stable from one iteration to the next. 
The inputs to RADEX are the gas density (\den), the kinetic temperature (\tkin), and the CO column density per unit line width
 (\ncolone/$dv$), which sets the optical depth scale. 

We used this code to compute CO intensities  for a large grid
in \tkin\ (10--3000 K), \den\ ($10^{2}$--$10^{8}$ \cmthree), and \ncolone/$dv$ ($10^{14.5}$--$10^{18.5}$ \cmtwo\  (km s$^{-1}$)$^{-1}$). From this grid, we generated likelihood distributions, adapting the method described in \citet{ward03} (also see Kamenetzky et al. 2011), for \tkin, \den, and \ncolone, by comparing the RADEX and observed line fluxes. 

To avoid any non-physical solutions we applied three priors in the likelihood analysis.  The first one limits the $^{12}$CO column density such that the  total mass of the molecular gas producing the CO lines cannot exceed the dynamical mass of the
galaxy according to the following relation:
\begin{equation}
\frac{N_\mathrm{CO}}{dv} < \frac{M_\mathrm{{dyn}}x_\mathrm{CO}}{\mu m_\mathrm{H_{2}}}\frac{1}{A\Delta V} 
= 1.6 \times 10^{18}\,\,\mathrm{cm}^{-2}\, (\mathrm{km}\,\mathrm{s}^{-1})^{-1}
\end{equation}
where the dynamical mass of the disk $M_\mathrm{dyn} = 1 \times 10^{10}$ \msun\ (SC97), the abundance of CO relative to H$_{2}$, $x_{\mathrm{CO}} = 3 \times 10^{-4}$, $A$ is the area in $\mathrm{cm}^{2}$ covered by the source and is given by A = (Angular diameter distance)$^{2}$$ \times \Omega_{s}$, 
and $\mu = 1.4$ is the mean molecular weight in units of $m_\mathrm{H}$. The source size, $\Omega_{s}$, is different for the high-J and the low-J lines; the values we used are described later in this section. To get the total CO column density, the \ncolone/$dv$ should be multiplied by the line velocity width ($\Delta V$). We use $\Delta V = 350$ \kms\ for all the FTS lines, which comes from the line fitting of J = 12-11 (which lies at a frequency where the spectral lines are resolved). 
The second prior limits the column length to be less than that of the entire molecular region according to

\begin{equation}
\frac{N(\mathrm{CO})}{n(\mathrm{H_{2}})x_{\mathrm{CO}}} < 2\, R_\mathrm{d}  \sim 400\, \mathrm{pc}
\end{equation}

where $R_\mathrm{d} \sim 200\, \mathrm{pc}$ is the radius of the emitting region. This prior will limit the density at the lower end and column density at the higher end.

The third prior is the absolute surface brightness of the CO lines that is set by the choice of our source size and an area filling factor of unity. This prior will effect the column density such that it will systematically shift to higher values for smaller assumed source sizes. 

We first modeled the J = 4-3 to J = 13-12 transitions and found that the highest likelihood model, with a temperature of $\sim 1350$ K, provides a good fit to all the transitions with the exception of J = 4-3. The model fit to the observed line temperatures is shown in Figure \ref{cosed} (right panel). The mid-J to high-J lines are tracing very warm molecular gas. The predicted Rayleigh-Jeans temperature of the J = 4-3 line is underestimated with respect to the observed temperatures by this warm component, implying a second temperature component. We then modeled the low-J lines from J = 1-0 to J = 4-3, after subtracting the contribution from the warm component, and found that these lines are fit well by a cold temperature component of $\sim 50K$ as shown by the dashed line in the right panel of Figure \ref{cosed}. The total of the two is shown by the dot-dashed line.

The likelihood distributions of \tkin, \den, \ncolone\  and thermal gas pressure (\tkin\ $\times$ \den) are shown in Figure \ref{co} for the warm (red) and cold (blue) components. The likelihoods for the warm component are narrow and provide very tight constraints on these parameters. Note that neither priors in Equation (3) and (4) were relevant for modeling the warm CO. In fact, only the third prior is important; the absolute line temperature set by the source size ($\Omega_{s} = 4.1 \times 10^{-11}\, \mathrm{sr}$) calculated in section \ref{dustext}. This source size was calculated assuming an optically thick emission, which is a reasonable value to use in absence of spatially resolved CO maps at these high frequencies. As mentioned previously, any change in the source size will affect the likelihood distributions, particularly the column density, which will shift systematically to lower or higher values. The effect on density and temperature likelihood is minimal. The optical depths of mid-J to high-J transitions range from 3.3 to 0.1, peaking at J = 6-5. 
The likelihoods for the cold component are also very well constrained. For the modeling of the cold component we used the source size of $6.4 \times 10^{-11}\, \mathrm{sr}$ and line velocity width of 500 \kms\ as measured by SC97 for the CO J = 1-0 line. In this case also the third prior is more important. The mass prior in Equation (3) does limit the column density at the higher end but just barely. The low-J transitions are optically thick.

The highest likelihood values of \tkin, \den, \ncolone, pressure, mass, and luminosity are listed in Table 3 for the two components. Their pressure likelihoods do not overlap implying that these two components are not co-spatial. In addition, the observed velocity widths are also significantly different between these components. Even though the mass of the warm gas is only about 10\% of the mass of the cold gas, the luminosity and hence the cooling is still dominated by the warm high-J CO lines. 

The molecular gas traced by CO in Arp~220 is similar to M82, where P10 also found warm and cold temperature components. However the temperature of these components were smaller in M82 by a factor of three and two for the warm and cold components, respectively. This is not unexpected given the higher FIR luminosity of Arp~220.

The CO emission that we observe, in both the warm and cold components, arises in fairly low density ($n \sim 10^3$ \cmthree). Undoubtedly denser gas components are present; for example, \citet{greve09} argue for gas with $n \sim 10^6$ \cmthree\ and $T\sim 50 - 120$~K based on their observations of CS lines. That we find no evidence for such a component indicates only that the contribution from this gas to the observed CO emission is small. 

\subsection{Comparison with H$_{2}$ Observations}
 \citet{higdon06} conducted a survey with the \textit{Spitzer Space Telescope} of warm molecular gas in nearby ULIRGs (including Arp~220) by measuring the optically thin rotational lines of \htwo. They list line intensity measurements for the S(1), S(2), S(3) and S(7) lines, and provide upper limits on the S(0) line. We applied a dust correction\footnote{We found the extinction correction factor, $I_{0}/I \sim 3.7$, for the S(1), S(2) and S(3) transitions, and $\sim 5.3$ for the S(7) transition.} to these line intensities using the mixed dust model and the mid infra-red extinction curve of \citet{krugel94} (also see \citet{haas01}), and used them to calculate the excitation temperature following the method described in sections 3 and 4 of \citet{higdon06}. We found the excitation temperature to be $\sim 1380$ K for the warm \htwo\ gas using the S(3) and S(7) lines, in excellent agreement with our warm CO temperature. We note that without making the extinction correction for \htwo\ and CO, the derived gas temperatures are grossly underestimated: $\sim 260$~K and $\sim 400$~K respectively. We estimated a total luminosity from the observed \htwo\ lines to be $L_{\mathrm{H}_2} \sim 2 \times 10^{8}\, L_{\odot}$,  and infer a warm \htwo\ mass of $M_{\mathrm{H}_2} \sim 3 \times 10^{8}\, M_{\odot}$. Both the luminosity and mass of warm \htwo\ are consistent with our warm CO measurements. This implies that the warm \htwo\ gas observed by \textit{Spitzer} is the same molecular gas traced by the high-J CO lines measured by \herschel. 
 
 The total \htwo\ cooling at 1350 K is about 20 $L_{\odot}/M_{\odot}$ using the theoretical \htwo\ cooling curve of \citet{bourlot99}. This is much larger than the total cooling from CO of $\sim 0.4\,L_{\odot}/M_{\odot}$ implying that \htwo\ is still the dominant coolant at high temperatures. The total \htwo\ cooling in Arp~220 is almost an order of magnitude large than M82 (P10); a consequence of the much higher temperature in Arp~220 than in M82 (1350 K vs 500 K). A cooling of 20 $L_{\odot}/M_{\odot}$ would imply that the warm molecular gas mass in Arp~220 would produce a luminosity of about $9 \times 10^{9}\, L_{\odot}$; a large number but still a small fraction ($5 \times 10^{-3}$) of the total FIR luminosity of Arp~220.
 \subsection{Heating Sources of the Warm CO Molecular Gas}
Here we consider four major sources of energy that could be responsible for producing the luminosity of the warm CO: heating by UV star-light in photo-dissociation regions, heating by X-rays in X-ray dominated regions, heating by cosmic rays (CR), and mechanical heating. 
\subsubsection{Photo-Dissociation Regions} 
Photo-dissociation regions are neutral regions present at the skins of molecular clouds that are illuminated by intense far-UV (FUV) radiation fields. This FUV radiation can strongly influence the gas chemistry in these regions and is also the most important source of heating. 

We explore a PDR model grid \citep{kaufman06, wolfire10} with predictions for CO line fluxes from J = 1-0 to J = 30-29, generated over a large range in density ( $10^{1} - 10^{7}$ \cmthree) and incident FUV flux, Log $G_{0}$ (-0.5 -- 6.5), where $G_{0}$ is the FUV flux in units of the local interstellar value also sometimes referred to as the Habing interstellar radiation field i.e., FUV flux = $1.3\times 10^{-4} \times G_{0}$ erg \cmtwo\ s$^{-1}$ sr$^{-1}$ \citep{wolfire10}. Also, these models assume that FIR flux = $2  \times$ FUV flux. We first compare the ratio of the CO J = 6-5 line flux to the FIR surface brightness in Figure \ref{pdrxdr} (top left). The red dotted contour lines are the ratio of the model CO (6-5) line flux to FIR surface brightness, the green dashed line is the observed ratio, and the black dashed lines show the $1-\sigma$ density limits from Table 3. The FIR flux is labeled on right hand axis of the left panel. The observed ratio can be produced within the acceptable density limits. However, the model FIR surface brightness is only $\sim 0.1$  erg \cmtwo\ s$^{-1}$ sr$^{-1}$ compared to the observed FIR surface brightness, which is about $\sim 70$ erg \cmtwo\ s$^{-1}$ sr$^{-1}$, 700 times larger. The PDR models fail to explain the observed FIR surface brightness. In principle, one can stack up multiple PDRs to reproduce the observed FIR. In our case, about 700 PDRs would have to be stacked to match the observed FIR surface brightness. However there is a limit to the number of PDRs that can be stacked, which is set by the characteristic column density of about $10^{22}$ \cmtwo. The observed column density for warm molecular gas is (from Table 3) $\sim 10^{23}$ \cmtwo\ (using $x_{\mathrm{CO}} = 3 \times 10^{-4}$), allowing a maximum of 10 PDRs for stacking, not sufficient to match the observed FIR surface brightness. Moreover, in Figure \ref{pdrxdr} (top right) we compare the line ratio of CO J = 12-11 to CO J = 5-4. The observed line ratio (green line) of 0.7 cannot be produced by PDR models within the accepted density limits. In fact, the densities required to match the observed ratios would have to be greater than $10^{5.5}$ \cmthree, completely excluded by our likelihood analysis. 

From these two comparisons, we conclude that PDRs cannot produce the observed warm CO luminosity and hence are not the dominant source of heating. They can only provide about 1.4\% of the observed FIR flux.
\subsubsection{X-ray Dominated Regions}\label{xdrmod}
The XDR modeling is done using an updated version of the code described in \citet[MHT]{maloney96II}. We assume a hard X-ray (1-100 keV) luminosity of $10^{44}$ erg s$^{-1}$, as would be expected for an object of Arp 220's luminosity if a large fraction of the bolometric luminosity ($\sim$ \lfir) were produced by an AGN. A power-law index $\alpha = 0.7$ is assumed. Unlike the models in MHT, we assume a sharp lower energy cutoff of the incident spectrum at 1 keV; this affects the results only at column densities $\lesssim 10^{22}$ cm$^{-2}$. We also assume an attenuating column of $10^{24}$ \cmtwo\ between the X-ray source and the region being modeled. This value is based on the inferred column densities in the cold component (modeled in Section \ref{comodel}). The physical and chemical state of the gas is calculated using an iterative scheme; radiative transfer of cooling radiation is handled with an escape probability treatment, including the effects of dust trapping of line photons.

Figure \ref{pdrxdr} shows XDR models for CO abundance (bottom left) and CO surface brightness (bottom right). In the left panel, red lines are the model CO abundances and in the right panel red lines are CO surface brightness; in both panels blue lines are the gas temperatures, and green lines are the expected/observed values. The black dashed rectangle shows the accepted range of the source size (radius from the center of the XDR) and limits on cloud density (Table 3). It extends up to about 240 pc as calculated from the dust SED in Section \ref{dustext}. The blue circle is the $1-\sigma$ temperature range from the CO non-LTE modeling (Table 3). Within this temperature range, the XDR models predict a CO abundance of $10^{-8}$ -- an extremely low value. At these X-ray fluxes and gas densities, the high ionization fractions result in rapid chemical destruction of the CO, and thus a low equilibrium abundance. In addition, the XDR models cannot match the observed CO surface brightness, as shown in the bottom right panel. In both cases, much higher densities $\gtrsim 10^{4.5}$ \cmthree\ will be required to match the observed values.  In fact, if we lower the X-ray luminosity to bring down the destruction rate of CO, the temperatures would go down, which is not consistent with the very warm temperatures that we derive from the CO emission. So under no condition can we produce the observed warm CO at such low densities. This is not evidence against an XDR in Arp~220, but shows that an XDR cannot be the source of the warm CO emission. Such low density molecular gas can still survive in the presence of an X-ray luminous AGN as long as it is shielded from it (from large dust column densities). 

We compare this case with the Cloverleaf quasar, which has a luminous XDR. A similar analysis was done by \citet{bradford09}, in which Z-spec measurements of high-J CO lines (up to J = 9-8)  combined with low-J lines from other instruments were used to model the CO. The density of the gas in Cloverleaf as derived from non-LTE modeling was much higher, $n\sim 10^{5}$ \cmthree\, and the temperature was much lower, $T \sim 100$~ K. The XDR models readily produce the observed CO surface brightness and abundance at the observed density, temperature and size scale of the emission. Thus, XDR excitation of CO is possible in galaxies, but we have ruled it out as a dominant heating mechanism for the warm molecular gas in Arp~220.
\subsubsection{Cosmic Rays}\label{crmod}
It is widely believed that the low-energy cosmic ray (CR) protons are efficient in ionizing and exciting the gas in the ISM. There is evidence of CR heating of molecular gas in our Galaxy (Goldsmith \& Langer 1978), and in external galaxies particularly the starbursts such as M82 (Suchkov et al. 1993) and NGC 253 (Bradford et al. 2003). 

\citet{meijerink11} [ME11] modeled the effect of CR heating on the chemistry of molecular gas in the ISM of Arp~220. They give predictions for column densities of CO and other species for CR rates ranging from the Galactic value of $5 \times 10^{-17}$ s$^{-1}$ to $5 \times 10^{-13}$ s$^{-1}$,  as expected in ULIRGs \citep{papad10II}. This range of CR rates are reasonable for a ULIRG when compared to the high CR rates of $\sim 10^3$ times the Galactic value, derived from the observations of gamma rays in M82 and NGC 253 \citep{acero09, abdo10}.  ME11 find that the total CO column density is insensitive to CR flux and therefore we cannot say much about the effect of CR on the warm CO from these models. However, the CR heating rate using the upper limit of the CR rate ($5 \times 10^{-13} \,\mathrm{s}^{-1}$) from ME11 is $\sim 7.5 \times 10^{-24}\, \mathrm{erg\, s}^{-1}$ per molecule, about two orders of magnitude lower than the energy budget from \htwo\ cooling of 20 $L_\odot/M_\odot$ ($\sim 1.7 \times 10^{-22}\, \mathrm{erg\, s}^{-1}$ per molecule). In other words, a CR rate of $\sim 10^{-11}\,\mathrm{s}^{-1}$ is required to satisfy the total \htwo\ cooling. This rate is about two orders of magnitude above the prediction for ULIRGs \citep{papad10II}. One would expect a distribution of  CR rates such that the tail of this distribution can provide the higher rates required to satisfy the energy budget. However, this would imply that we should also see a distribution in temperature of the warm CO gas, which is not reflected in the narrow temperature likelihood we obtained from the non-LTE modeling.  In summary, CR are not likely to have sufficient energy required to explain the observed warm molecular gas. 
\subsubsection{Non-Ionizing Energy Source}
None of the above sources can produce the observed warm CO. We believe that a non-ionizing source like mechanical energy is required to heat the molecular gas. Large mechanical energy is expected in a merging system like Arp~220. This is also supported by a velocity gradient ($v_\mathrm{rms} = N_{\mathrm{H}_{2}} \Delta V/n_{\mathrm{H_2}}$) as large as 20 \kms\ pc$^{-1}$ we find from our RADEX likelihood analysis. Since we do not know the local size scale corresponding to this velocity gradient the amount of mechanical energy cannot be determined. But we can obtain a rough estimate of energy from turbulent heating for a region containing 1 M$_{\odot}$ of gas of density $\sim 10^{3}$ \cmthree. The length scale corresponding to this volume is about $\sim 0.25\, \mathrm{pc}$. Over this length scale the turbulent velocity will be about 5 \kms\ (using the velocity gradient of 20 \kms\ pc$^{-1}$). Using this velocity in the expression of turbulent heating per unit mass from \citet{bradford05} we obtained 0.01 $L_\odot/M_\odot$, which is much lower than the observed \htwo\ cooling. Unlike M82 (P10), in which turbulent heating was likely the major source of mechanical energy, it is not the case in Arp~220. 

Finally, we consider the energy from supernovae and stellar winds. Using the observed supernova rate of $\nu_\mathrm{SN} = 4 \pm 2$ per year \citep{lonsdale06}  and ($E_\mathrm{SN} = 10^{51}$  erg) in the following expression from \citet{maloney99}
\begin{equation}
L_\mathrm{SN} \sim 3 \times 10^{43} (\frac{\nu_\mathrm{SN}}{1 \mathrm{yr}^{-1}})  (\frac{E_\mathrm{SN}}{10^{51}\,\mathrm{ergs}})\, \mathrm {erg}\,\mathrm{s}^{-1}, 
\end{equation}
we obtain total mechanical energy of $\sim 100\, L_\odot/M_\odot$ (using $M_\mathrm{H_{2}} \sim 3 \times 10^{8}\,M_\odot$) from supernovae in Arp~220. The energy output from stellar winds is similar to supernovae (McCray 1987) assuming a Salpeter initial mass function, giving a total of $\sim 200$  $L_\odot/M_\odot$. This is large enough such that a small portion of it will be sufficient to match the observed \htwo\ cooling. In conclusion, mechanical energy from stellar winds and supernova are strong candidates for heating of the warm molecular gas in Arp~220.
\section{HCN in Absorption: Infrared Pumping}
HCN can be a very good tracer of dense molecular gas \citep{gao04}: its rotational transitions have critical densities 100 -- 1000 times higher than CO. We detect very high-J HCN lines, from J = 12-11 to J = 17-16, in absorption, which have critical densities around $10^9 - 10^{10}$ \cmtwo. Their observed equivalent widths are listed in Table 1, with the exception of the J = 13-12 line, which is blended with strong CO and water transitions. Such high-J transitions of HCN have never been observed before in other galaxies, in emission or absorption. Their low-J counterparts from J = 1-0 to J = 4-3 have been observed in emission from ground-based observatories. The left panel of Figure \ref{hcncog} shows the observed HCN ladder. The transition from emission to absorption happens somewhere between J = 4-3 and J = 12-11.  

All the HCN absorption lines fall on the linear part of the curve of growth as shown in Figure \ref{hcncog}, and therefore their column densities can be measured directly from the observed equivalent width per the following relation \citep{spitzer68}
\begin{equation}\label{cogeq}
\frac{W_{\lambda}}{\lambda} = 8.85 \times 10^{-13} N_{j}\lambda f_{jk}
\end{equation}
where $W_{\lambda}$ is the equivalent width, $N_{j}$ is the column density in the jth level, and $f$ is the oscillator strength. The unit of the constant in the above equation is cm. The summed observed column density of HCN from the five high-J absorption lines is $\sim 5.4 \times 10^{14}$ \cmtwo.

For estimating the total column density and the physical conditions of the gas that this HCN is tracing, we again do a RADEX likelihood analysis, but this time we use the column densities of the absorption lines instead of line fluxes. Using column densities directly eliminates any dependence on the source size or the line velocity width. Figure \ref{hcnlike} shows the results of our likelihood analysis. The temperature and density are not well constrained, so we only provide lower limits for them in Table 3. However, the column density is extremely well constrained with a total column of $N_\mathrm{HCN} \sim 2 \times 10^{15}$ \cmtwo. The RADEX analysis suggests that the gas is warm, very dense and has a pressure of $\gtrsim 4 \times 10^9$ K \cmthree.

However, there are several lines of evidence that strongly argue that collisional excitation (as assumed in the RADEX models) is not the dominant mechanism for populating these levels:  (a) The v$_2$ = 1 level lies 1025~K above the ground. In order for it to be effectively populated by collisions, the gas kinetic temperature must be a substantial fraction of this value. In addition, the densities must be high, since the vibrational Einstein $A$-coefficients are large ($\sim1$ \kms). This will lead to thermalization of the rotational levels at some T comparable to 1025~K. However, the rotational temperature of the absorption lines is only $\sim 300$~K. (b) Transitions within the v$_2$ = 1 vibrational state were directly observed by \citet{salter08} using the Arecibo radio telescope. These also have a low rotational temperature that is very similar to the ground vibrational state J-levels seen in absorption. This again argues that the kinetic temperature cannot be large enough to collisionally populate v$_2$=1, and (c) From the observed lines within v$_2$ = 1 levels we can estimate the (v$_{2}=1 $, J=6) column density. This is 3 times larger than the (v$_2$=0, J=6) column predicted by the RADEX collisional excitation analysis, and is a very large fraction of the total v$_2$=0 HCN column derived from the RADEX results. This is unphysical. Therefore, we believe that the high-J lines are populated not by collisional excitation but instead by radiative pumping of infrared (IR) photons \citep{carroll81}. 

The IR pumping mechanism works as follows: Assuming that the collisional excitation is negligible, an HCN molecule in v=0, J =0 ground state can absorb a 14~\mic\ photon from the dust continuum and be excited to v$_2$=1, J=1 state (as per the selection rule of $\Delta J = \pm 1$). This HCN molecule can decay back either to v=0, J =0 or v=0, J = 2 with equal probabilities. If the IR pumping rate is large enough, then the molecule in J = 2 ground state can absorb another 14~\mic\ photon before decaying to J = 1 and J = 0, and can be excited to v$_2$=1, J = 3, which can decay into J = 4 ground state, thereby raising the excitation of the ground state rotational levels. This mechanism can be efficient in populating the low-J HCN lines. But for an intense radiation field, very high-J ground state transitions can also be populated. For IR pumping of the vibrational states to significantly affect the population of the ground state rotational J+1 level, the pumping rate out of the ground state must exceed the spontaneous radiative decay rate out of the J+1 level. This gives the following criteria (Eq. 5 of \citet{carroll81}) for the temperature of the radiation field

\begin{equation}
\frac{f} {e^{\mathrm{h\nu}/\mathrm{kT_d}} - 1} > \frac{A_{J+1,J}}{A_{v,v-1}},
\end{equation}
\\
where $f$ is the dilution factor to account for both the geometric dilution and optical depth effects, $T_d$ is the temperature of the radiation field, and A's are the Einstein coefficient for spontaneous decay. The $h\nu/k$ is 1025 K for the v$_2$=1, and substantially larger for the other vibrational levels, so it is unlikely that any higher levels will contribute to the IR pumping. 

To populate J = 17-16, the above criteria will require the temperature of the radiation field to be $T_D \gtrsim 350 K$, assuming $f =1$ for a blackbody radiation field. For $f < 1$ this temperature must be even higher. The implication here is that the gas observed in absorption in HCN needs to see a $> 350$~K radiation field. However, this does not mean that we will find evidence for this thermal component in our dust continuum data. The line of sight to this continuum component is likely to be highly obscured.

There is direct evidence for IR pumping via the 14~\mic\ the HCN absorption feature in the Arp~220 spectrum obtained with \textit{Spitzer} \citep{lahuis07}.
The column density of this 14~\mic\ absorption is a factor of $\sim 15$ larger (with 30\% uncertainty) than the HCN column we obtain from the absorption lines. It is difficult to determine if the gas traced by the 14~\mic\ bending-mode absorption feature is the same component seen in the pure rotation absorption line, especially given the large optical depths in Arp~220. Additional evidence for strong IR pumping comes from the observation of the HNC J = 3-2 line, which is overluminous in Arp~220; it is brighter than the HCN J = 3-2 line by a factor of $\sim 2$ \citep{aalto07}. Vibrationally excited HCN lines have been also detected in other luminous infrared galaxies such as NGC 4418 \citep{sakamoto10}. There is also evidence for IR pumping of higher-J HCN lines in the host galaxy of the $z = 3.91$ quasar APM 08279+5255 \citep{riechers10,weiss07}. This is particularly relevant to this work as it is argued by \citet{downes07} that the dust emission in the western nucleus of Arp 220 is strikingly similar to this quasar. If the excitation of the high-J HCN lines in Arp~220 is dominated by IR pumping rather than collisions, then the HCN is not necessarily tracing dense gas. 

In summary, we propose that the populations of the high-J levels of HCN seen in absorption in Arp~220 are produced dominantly by the effects of radiative pumping, along with the additional requirement of intense radiation field with temperature of $> 350$ K, while the low-J levels of HCN (J = 1-0 to J = 4-3 observed from the ground) are affected by both collisions and radiative pumping, thereby increasing their column density. To model the HCN in Arp~220 we will need to include effects from both IR pumping and collisions in our RADEX code. This is beyond the scope of this paper.  
\section{Atomic Lines: [\ion{C}{1}] and [\ion{N}{2}]}\label{atomic}
As mentioned in Section \ref{lineid}, we detect two [\ion{C}{1}] lines and one [\ion{N}{2}] line in emission. The [\ion{C}{1}] lines are very difficult to observe from the ground and so far they have been detected in only a handful of galaxies, including M82 (P10). The ratio of line strengths (in Jy \kms) of [\ion{C}{1}] (2-1) to [\ion{C}{1}] (1-0) in Arp~220 after correcting for dust extinction is $1.3 \pm 0.3$, consistent with unity within the error bars. This suggests that these lines are optically thick, which is also supported by the excitation temperature of 26 K derived from their line temperature ratio \citep{stutzki97}, which is too low compared to the kinetic temperature of the cold molecular gas (50~K). For comparison, in M82, using FTS measurements of these lines (P10) we found this ratio to be $2.1 \pm 0.2$. To calculate the [\ion{C}{1}] column density in Arp~220 we use the escape probability radiative transfer models from Figure 2 of \citet{stutzki97}. For the observed line temperature (in K \kms) ratio of [\ion{C}{1}] (2-1) to [\ion{C}{1}] (1-0) of  $\sim 0.5$, and the brightness temperature of $\sim 13.5$~K of [\ion{C}{1}] (1-0) line (assuming 500 \kms\ line width), we estimate a column density of $N_\mathrm{C}  \sim 10^{20}$ \cmtwo. The low excitation temperature of \ion{C}{1} suggests that it is associated with the cold molecular gas traced by low-J CO lines with the ratio of $N_\mathrm{C}/N_\mathrm{CO} \sim 1$. 

The ratio $N_\mathrm{C}/N_\mathrm{CO}$ is much higher in Arp~220 compared to the Galactic PDRs, e.g, this ratio is around 0.17 for the Orion bar (Tauber et al. 1995).  Previous studies of atomic carbon in galaxies \citep[e.g.,][]{wilson97,israel05} suggest that [\ion{C}{1}] emission gets stronger for more luminous systems with starbursts and AGN. For comparison, the ratio of $N_\mathrm{[C]}/N_\mathrm{CO}$ in M82 and the Cloverleaf Quasar are 0.5 and 1, respectively. \citet{keene96} found one-to-one correlation between [\ion{C}{1}] and $^{13}$CO  line strengths for Galactic PDRs. They argue that simple PDR models in which the cloud has a plane parallel geometry illuminated on one side by UV field cannot produce this correlation. In these simplified models, [\ion{C}{1}] forms a narrow layer between [\ion{C}{2}] on the outside and CO deeper in the molecular cloud, with column density of about $5 \times 10^{17}$ \cmtwo. The column density and line strength of [\ion{C}{1}] (1-0) line in these models are relatively insensitive to physical parameters such as temperature, density and the radiation field. The observations of some Galactic PDRs have shown that the [\ion{C}{1}] can extend far deeper into the molecular clouds, much further than predicted by simple PDR models. \citet{israel05} found from the survey of atomic carbon in nearby luminous galaxies, that the correlation between [\ion{C}{1}] and $^{13}$CO still holds but it is not one to one anymore. Instead [\ion{C}{1}] emission gets stronger compared to CO for more luminous systems. This is shown in their Figure 2 in which the ratio of [\ion{C}{1}] (1-0) and $^{13}$CO (2-1) is plotted as a function of [\ion{C}{1}] luminosity. On this plot, Arp~220 falls on the upper right corner (with coordinates $\sim (3.2, 4.5)$),  consistent with the correlation.  

The extinction corrected luminosity of the [\ion{N}{2}] 205 \mic\ line is $\sim 2.96 \times 10^{7}$ $L_{\odot}$, which is 10 times higher than that of M82 (P10). But compared to their respective FIR luminosities, the  $L_\mathrm{[NII]}/L_\mathrm{FIR}$ ratio of $\sim 1.7 \times 10^{-5}$ in Arp~220 is lower than M82 by a factor of 3. The deficit of [\ion{N}{2}] 205 \mic\ line relative to \lfir\ in Arp~220 is surprising because [\ion{N}{2}] is produced in the \ion{H}{2} regions and not in the photo-dissociation regions, where the [\ion{C}{2}] 158 \mic\ deficiency is generally found in ULIRGs \citep{malhotra97}. However, the [\ion{N}{2}] line deficit we find here is consistent with the recent FIR line deficit results from \herschel -- PACS presented in \citet{gracia11}. These authors found that the [\ion{N}{2}] 122 $\mu$m line shows deficits similar to the [\ion{C}{2}] 158 \mic\ line deficit relative to \lfir\ for a large sample of galaxies. They claim that this is the case with all FIR lines but their case for an [OIII] line deficit is weak compared to [\ion{N}{2}] line deficit. Their modeling results suggest that the line deficit in \ion{H}{2} regions is the result of a higher ionization parameter in ULIRGs compared to normal or starburst galaxies. A higher ionization parameter will increase the dust absorption of the UV photons, which will decrease the fraction of photons available for ionizing the neutral gas thereby decreasing the fluxes of [\ion{N}{2}] and other FIR lines relative to the continuum \citep{abel09}.  

The line ratio of [\ion{N}{2}] 122 $\mu$m to 205 $\mu$m is an excellent density probe of the ionized gas. However, in Arp~220 the [\ion{N}{2}] 122~$\mu$m appears in absorption in the ISO spectrum \citep{brauher08}, so we cannot use this line to estimate the electron density. Instead we compare the line strength of the [\ion{N}{2}] 205 $\mu$m line to the [\ion{C}{2}] 158 $\mu$m line measured by ISO \citep{brauher08}. The line ratio of [\ion{C}{2}]/[\ion{N}{2}] gives an estimate of the fraction of [\ion{C}{2}] in the ionized medium compared to the PDRs (e.g., \citet{oberst06}). We find this line ratio (corrected for dust extinction) to be $17 \pm 3$. We compare this ratio to Figure 2 of \citet{oberst06}, in which they plot the model line intensity ratio as a function of ionized gas density. Because we cannot estimate the ionized gas density in Arp~220, we consider the minimum and maximum values of this line ratio over the entire density range. These model line ratios range from 2.5 -- 4.3. Comparing this model line ratio to our observed value of 17, we constrain the contribution from \ion{H}{2} regions to the observed [\ion{C}{2}] 158 $\mu$m emission to be between 15\% and 26\%. This suggests that the majority of [\ion{C}{2}] arises in the PDRs, as concluded in the previous works \citep[e.g.,][]{ madden93, sauty98}.  
\section{The Molecular Outflow and Water Chemistry}
\subsection{Molecular Outflow: Observations of P Cygni Profiles}
In the standard scenario of galaxy evolution, some of the fundamental open questions are: How do feedback mechanisms regulate star formation and supermassive black hole growth to produce the observed $M-\sigma$ relation between stellar velocity dispersions and black hole masses \citep{laura00, gebhardt00}?  Can the energy injection from starbursts or AGN quench star formation by expelling the gas from the host galaxy? Is this the process by which massive gas-rich galaxies deplete their interstellar media to become gas-poor early-type galaxies?  Direct observational evidence of massive molecular outflows associated with AGN/starbursts can significantly impact our understanding of the connection between feedback and galaxy evolution. 

Observations of nearby galaxies with AGN have shown outflows but generally only in the ionized gas component, which is a minor fraction of the gas mass in
the host galaxy. Direct observational evidence of massive molecular outflows was recently discovered in two galaxies using the IRAM PdBI and \herschel\ PACS: Mrk~231 \citep{fischer2010, feruglio2010} and NGC~1266 \citep{alatolo11}. In Mrk~231, \citet{fischer2010} detected a massive outflow with velocity of $\sim 1400$ \kms\ using the P Cygni profile seen in the OH lines. Independently, \citet{feruglio2010} also detected the same outflow using the CO (1-0) line, and derived an outflow rate of 700~M$_\odot$/year, which will deplete the gas reservoir in $10^{7}$ yrs. In NGC 1266, a nearby field S0 galaxy, \citet{alatolo11} discovered a massive molecular outflow originating from the nucleus. They found the outflow velocity of 400 \kms\ and outflow rate of  13~M$_\odot$/yr with a depletion time scale of 85 Myr. They conclude that the outflow has to be drive by an AGN in absence of sufficient star-formation.

A P Cygni line profile (redshifted emission peak with corresponding blue-shifted absorption)  is a strong signature of outflow activity. We detect P Cygni profiles in the OH$^{+}$, \water$^{+}$ and HF lines as shown in Figure \ref{pcygni},  suggesting molecular outflow. Since our P Cygni line profile is not resolved we can only put an upper limit on the outflow velocity of about $250$ \kms. 
There is independent evidence for molecular outflow from ground-based observations by \citet{sakamoto09}, in which P~Cygni profiles were detected in HCO$^+$ lines. They found that the outflow has a velocity of $100 - 200$ \kms\ and is confined within a radius of 50 pc. The total HCO$^{+}$ column density was about $10^{15}$ \cmtwo.  The outflow in HCO$^{+}$ is very likely the same outflow that we detect in OH$^{+}$ and \water$^{+}$. Our upper limit of 250 \kms\ on the outflow velocity is a limit on the line-of-sight velocity. It is likely that the outflow has a wide angle and is not in the form of a narrow jet. In this situation, the maximum outflow velocity will be higher than 250 \kms\ depending on the inclination of the disk. Using the inclination angle of 45$^{\circ}$ from \citet{scoville97} and assuming that the outflow is orthogonal to the disk, the maximum outflow velocity is $\sim 350$ \kms. As this is lower than the escape velocity of the nuclear region of Arp 220 ($\sim 420$ \kms\ at a nuclear radius of 250 pc), we believe the outflow is gravitationally bound to the nuclear region.
\subsection{Importance of \water$^{+}$ Detection}
The detection of \water$^{+}$ by \herschel\ in the Galactic and extragalactic ISM was very surprising. This is because \water$^{+}$ is highly reactive and is expected to rapidly react with \htwo\ to form H$_{3}$O$^{+}$. The absence of H$_{3}$O$^{+}$ and a presence of strong absorption of  \water$^{+}$  requires a gas with a low molecular fraction, which can be the result of high cosmic ray or X-ray fluxes. The detection of  \water$^{+}$ and its implication on the physical conditions of the molecular gas is reviewed by \citet{vandish11}. The presence of strong absorption features in both \water$^{+}$ and OH$^{+}$ also suggest that the production of water in the gas associated with the outflow is dominated by low temperature ion-neutral chemistry. At high temperatures (T $\gtrsim 500$ K), the water production is dominated by neutral-neutral chemical reaction (no production of  \water$^{+}$ and OH$^{+}$), which increases the abundance of water significantly.    

\subsection{Origin of the Ionic Molecular Species: Evidence for AGN in Arp~220}
The presence of a hidden AGN in Arp~220 has long been debated. The X-ray observations from ROSAT \citep{heckman96} and \textit{Chandra} \citep{clements02} reveal a relatively weak X-ray source with a luminosity of $4 \times 10^{40}\, \mathrm{ergs\,s}^{-1}$. Neither study can rule out a hidden AGN that is obscured by the large column of N(\htwo) $\sim 10^{25}$ \cmtwo\, found in this study and in previous observations \citep[e.g.,][]{alfonso04}. In addition, \citet{haas01} found an exceptionally low ratio of 7.7 \mic\ PAH flux to submm continuum for Arp~220 and argue that a hidden AGN is required to power much of the luminosity of Arp~220. \citet{downes07} found evidence for a hot, compact, and optically thick dust ring in the western nucleus from 1.3 mm IRAM data, which they interpreted as being heated by an AGN accretion disk. More recently, \citet{engel11} found that this putative AGN's position is coincident with the stellar and CO (2-1) kinematic center, suggesting that the stars in the western nucleus are gravitationally bound to either a supermassive black hole or an extremely dense, young nuclear starburst.

Here we consider both CR and XDR models to determine which one of these sources can match the observed column densities of the ions involved in the chemical network producing water. The absorption lines of these ionic species are saturated; their equivalent widths are not linear on the curve of growth. They appear saturated because of the lower resolution of our FTS spectrum. Hence, we can only determine the lower limits on their column densities from their equivalent widths (listed in Table 2) using the relation in Equation \ref{cogeq}. We obtain $N$(OH$^{+}) \gtrsim  10^{15.6}$ \cmtwo, $N$(\water$^{+}) \gtrsim 10^{15.1}$ \cmtwo, and $N$(\water) $\gtrsim 14.3$ \cmtwo. These lower limits are calculated from the ground state transitions of these molecules, which contain a majority of the population. The H$_3$O$^{+}$ line is absent in our spectrum so we calculate a 3$-\sigma$ upper limit of N(H$_3$O$^{+}) \lesssim 10^{14.2}$ \cmtwo.  We compared the lower limits, $N$(OH$^{+}$) and $N$(\water$^{+}$), with the CR models of \citet{meijerink11} (see their Figure 7). They explored a wide range of the CR ionization rates as discussed in \ref{crmod}. We find that none of their models (with low and high densities; low and high $G_{0}$ fields) can produce the observed column densities for any value of CR ionization rates. These CR models underestimate these column densities of OH$^{+}$ and \water$^{+}$ by almost an order of magnitude, and overestimate the H$_3$O$^{+}$ column. 

We now consider XDR models, which were explained in \ref{xdrmod}. In Figures \ref{xdr} a -- e, we present predictions for the column density over the $n_\mathrm{H}$ and $N_\mathrm{H}$ grid for all the ions including HCO$^{+}$ that are involved in the outflow. The allowed regions of columns densities are shown by shaded area. This model was run for a source size of 50 pc (estimated for HCO$^{+}$ by \citet{sakamoto09}). In Figure \ref{xdr}f, total column densities of each species is shown as a function of total hydrogen column and the observed lower limits are marked by dashed lines\footnote{For H$_3$O$^+$ and HCO$^+$, the dashed line represents an upper limit and a total column density respectively.}. Compared to CR, the XDR models can readily produce the column densities of all these species. In order to produce the observed columns, the luminosity of the XDR has to be $\sim 10^{44}$ ergs s$^{-1}$.

In figures \ref{xdr} a -- e, we note that OH$^{+}$ column density cannot be produced for $n_\mathrm{H} \gtrsim 10^{5}$ \cmthree. This gives us an upper limit on the density of the gas associated with the outflow. This limit will change with source size. We tested this by running XDR models for 100~pc and 240~pc source sizes. With a larger source size the density limit after which OH$^{+}$ column density cannot be matched goes down. It is also possible that there is a density gradient in the outflow, in which case the HCO$^{+}$ emission could be dominated by a denser component. There is support for this argument in Figure \ref{xdr}f, where the column densities of all the ions except HCO$^{+}$ correspond to the hydrogen column density of $\sim 2 \times 10^{23}$ \cmtwo. The HCO$^{+}$ column density corresponds to a much higher column density of $\sim 6 \times 10^{23}$ \cmtwo\ suggesting that HCO$^{+}$ could be coming from denser part ($n_\mathrm{H} \sim 10^{6}$ \cmthree) of the outflow. 

We use the hydrogen column density of $\sim 2 \times 10^{23}$ at which the column densities of all the ions (except HCO$^{+}$) are satisfied, to estimate the lower limit on the amount of mass associated with the bound outflow. We find this lower limit to be $M_\mathrm{outflow} \gtrsim 2 \times 10^{7}$ M$_{\odot}$, using a source size of 50 pc and conservatively assuming that the outflow only covers the continuum source. 

One of the interesting things about the Arp~220 outflow is that it appears to be very different from Mrk~231 and NGC~1266. This is because the mass associated with the outflow in Arp~220 is large but it will not escape the nuclear region of this galaxy.  Does this mean that the standard scenario of AGN/starbursts driving the molecular gas out of the galaxy does not apply here or is Arp~220 in a different evolutionary stage compared to Mrk~231 such that at some time in the future this gas will be expelled out of the galaxy?

In conclusion, the CR models not only fail to produce the observed column densities of \water$^+$ and OH$^+$ by almost an order of magnitude but they also over estimate H$_3$O$^+$ by a large amount. An XDR with luminosity of  $10^{44}$ ergs s$^{-1}$ can easily match the observations of all the above ionic species, strongly favoring the existence of an AGN in Arp~220.

\section{Discussion: Is Arp~220 a Good Template for High-z Submillimeter Galaxies?}
Arp~220's spectrum is commonly used as a template for high-$z$ submm galaxies because it is bright, and therefore has been extensively studied across the electromagnetic spectrum. We discuss here in the light of new results from this study on dust and molecular gas, whether Arp~220 is still a good template. We compare some characteristics of Arp~220 with several high-$z$ SMGs for which the molecular and continuum data were recently published. There are four high-$z$ sources, SMMJ2135-0102 \citep{danielson11}, GN20 \citep{carilli10}, SMM18423+5938 \citep{lestrade10}, and SMM10571+5730 \citep{scott11}, with $z$ ranging from 2.3 -- 4.0, for which submm spectroscopic data was obtained using several ground-based facilities (Z-Spec, IRAM-PdBI, APEX, SMA, CARMA, etc.).  Several CO transitions from low-J to mid-J were observed for all four systems. In all cases, the CO SLED peaked around CO (6-5) and then turned over, similar to Arp~220 and M82. Does this mean that the CO excitation in these high-$z$ systems is dominated by starbursts and not AGN? For the latter, the CO SLED is not expected to turn over around mid-J levels \citep{werf09, bradford09, pwerf2010}. More recently, EVLA observations of five high-$z$ quasar host galaxies \citep{riechers11, ivison11} indicate that SMGs with AGN show no cold gas component; i.e., the CO J = 1-0 line strengths in these five systems are consistent with warm, highly excited gas that is also seen in the higher-J CO lines. They conclude that the gas-rich quasars and AGN-free SMGs represent different stages in the early evolution of massive galaxies. It is complicated to compare this scenario directly with our findings in Arp~220, in which the CO SLED shape is consistent with gas that is excited by star-formation but observations of other molecules also indicate the presence of a luminous AGN. \citet{lupu10} observed five submm-bright lensed sources recently detected by the Herschel Astrophysical Terahertz Large Area Survey (H-ATLAS) using Z-Spec. Up to 3 mid-J CO transitions were detected in these systems, not enough to compare the shape of the SLED to Arp~220. 

We compare the ratio of $L_\mathrm{CO, J = 6-5}/L_\mathrm{FIR}$ in Arp~220 with a sample (discussed above) of 9 high-$z$ sources (using individual CO line luminosities) and find that this ratio does not vary much: it is $\sim$ a few $\times 10^{-5}$. In six \citep{scott11,lupu10} out of nine galaxies, the modeling of CO was similar to this work, but the results from this modeling have large error bars because fewer lines were available. This makes it hard to conclude if all these high-$z$ systems also have warm molecular gas as observed in Arp~220. The CO analysis by \citet{scott11}, for which six CO transitions were available, indicates that HLSW-01 has significant amount of warm molecular gas; the CO SLED was fit well by single temperature component with \tkin\ ranging from 86 -- 230~K, with no evidence for a cold temperature component. The data in this system are consistent with no cold gas component. In addition, we compare the ratio of  $L_\mathrm{CI}/L_\mathrm{FIR} \sim 6.7 \times 10^{-6}$ in Arp~220 with a recent survey of CI in high-$z$ galaxies \citep{walter11}. This study found the ratio to be $\sim 7.7 \times 10^{-6}$ from a sample of high-redshift galaxies. The two ratios are consistent with each other and are roughly equal to the luminosity of a single CO line in Arp~220, implying that CI is not an important coolant compared to CO in Arp~220 and the high-$z$ sample.

In the six \citep{scott11,lupu10} out of nine galaxies, the modeling of dust was similar to this work; i.e., they were modeled using Equation 1. All these six high-$z$ systems show warm dust emission with $T_\mathrm{dust}$ varying from 55~K -- 90~K, and high optical depths of $\tau_{d} \sim 3.5 - 5$ at 100 \mic. However, for most of these galaxies $\beta$ was fixed at 2.0; since $\beta$ is correlated with $T_\mathrm{dust}$ (as shown in Figure \ref{dust}),  the dust temperatures could be lower. Both the dust temperature and high optical depth in Arp~220 are consistent with the range of temperature and optical depth estimated for the high-$z$ systems. Due to large optical depths, dust extinction correction should be considered for the CO lines observed in these SMGs. As demonstrated in this work, the kinetic temperature of the warm molecular gas is grossly underestimated (see section 5.2) if the CO line fluxes are not corrected for extinction. We note that this high-$z$ sample is small and suffers from a selection effect which favors the detection of warm galaxies for a given dust mass. 

In summary, even though FTS observations of Arp~220 indicate that it is a peculiar ULIRG with unusually large dust optical depths, massive molecular outflow, strong IR pumping and warm molecular gas, the global properties such as CO line to FIR continuum ratios, dust temperatures and shape of the CO SLED seems to be consistent with a sample of high-$z$ SMGs. Currently, the SMG sample is small and the S/N of the line observations is not as good as for Arp~220. It will be important to revisit this comparison as more molecular line observations of dusty, luminous, high-$z$ become available.

\section{Conclusions}

In this paper, we presented a complete spectral view of Arp~220 from 190 -- 670 \mic\ using the SPIRE-FTS; most of this wavelength region has not been accessible prior to \herschel. The continuous wavelength coverage provided by the FTS eliminates any systematic uncertainties due to cross-calibration issues arising from using multiple instruments observing over different wavelength windows, as has been the case for ground-based observations of nearby galaxies for a long time. The wavelength region covered by the FTS contains the higher-J rotational levels of CO, HCN and other molecules, which were crucial for carrying out a comprehensive characterization of the molecular gas. The continuous wavelength coverage also facilitated detections of rare molecular and atomic species that a targeted narrow band search would not have achieved. In addition to CO and water species, we report unexpected detections of very high-J absorption lines of HCN. The spectrum also show strong absorption in OH$^{+}$, \water$^{+}$, CH$^+$, HF, and several nitrogen hydrides. The two emission features of \water$^{+}$ at 742 and 746 GHz have been seen for the first time. We also detected \ion{C}{1} and [\ion{N}{2}] lines. These atomic and molecular species are very difficult to observe from the ground, and in some cases completely inaccessible. In addition to line detections, the FTS also provided a good measurement of the continuum, which is invaluable in characterizing dust in the nearby starbursts and ULIRGs. Hence, FTS instrument is playing a very important role in providing information at wavelengths that will be complementary to ALMA and other ground- and space-based telescopes. 

Below we list the conclusions for Arp~220, which are drawn from the analysis of the FTS spectrum:
\\
\\
(a) Modeling the continuum revealed dust that is warm with $T = 66$~K, and has an unusually high optical depth, with $\tau_{d,100\mu\mathrm{m}} \sim 5$. We derive a dust mass of about $10^{8}$ $M_\odot$ and a high total hydrogen column density of $\sim 10^{25}$ \cmtwo\, with an uncertainty of a factor of $\sim 3$, completely dominated by the uncertainty in the dust cross section.
\\
\\
(b) Warm Molecular gas: The extinction corrected observed CO luminosity is dominated by the mid-J to high-J lines, peaking around CO(6-5). The non-LTE radiative transfer modeling strongly indicates that the mid-J to high-J lines are tracing warm molecular gas with \tkin$ \sim 1350$~K. The low-J transitions are tracing cold gas with \tkin$ \sim 50$~K. The inferred temperature for the warm and cold components are much lower if CO line fluxes are not corrected for dust extinction. These two components are not in pressure equilibrium and the observations show that their line widths are different by a factor of $\sim 1.5$. The mass of the warm gas is about 10\% of the cold molecular gas but dominates the luminosity as well as the cooling over the cold CO. The ratio $L_\mathrm{CO}/L_\mathrm{FIR} \sim 10^{-4}$, where $L_\mathrm{CO}$ is the total CO luminosity. The temperature of the warm molecular gas is in excellent agreement with the temperature derived from \htwo\ rotational lines observed from \textit{Spitzer}, implying that CO is still a good tracer of \htwo\ at these high temperatures. At 1350~K, \htwo\ dominates the cooling of the ISM over CO. We also conclude that the contribution of the dense gas to the observed CO emission is small.

We have ruled out PDRs, XDRs and cosmic rays as possible sources of this warm molecular gas within the context of models we compared our data to. The mechanical energy from supernova and stellar winds can satisfy the energy budget required to heat this gas but we still do not know the exact mechanism that heats this gas. Such warm molecular gas has been confirmed in only two galaxies so far, M82 \citep{panuzzo2010} and Arp~220, as part of the submm wavelength region covering high-J transitions of CO was made accessible by SPIRE. In both cases some type of non-ionizing source is required to heat this gas. As FTS data for more galaxies become available, this issue will be investigated further. 
\\
\\
(c) The very high-J lines of HCN appear in absorption. The transitions from emission to absorption takes places somewhere between J = 4-5 and J = 12-11. These high-J lines are populated via IR pumping of photons at 14 \mic. The condition for IR pumping to populate J = 17-16 level requires an intense radiation field with $T > 350$~K.
\\
\\
(d) Massive molecular outflow: The signature of a molecular outflow is seen in the P Cygni profiles of OH$^+$, \water$^+$ and \water; major molecules involved in the ion-neutral chemistry producing water in the ISM. The outflow has a mass of $\sim 10^7$ M$_\odot$ and velocity $\lesssim 250$ \kms. It is massive but bound because its velocity is less than the escape velocity of the Arp~220 nuclei. The upcoming Atacama Large Millimeter Array (ALMA) will have the capability to map this outflow. A high resolution follow-up from HIFI and ALMA is required to accurately characterize this outflow. So far, massive molecular outflows have only been detected in Mrk~231 and NGC~1266. \\
\\
(e) The existence of an AGN in Arp~220 has long been debated. We found significant evidence for an AGN in Arp~220. Our modeling shows that the large observed column densities in OH$^+$, \water$^+$ and \water\ can only be produced by a luminous  XDR with $L_{X} = 10^{44}$ ergs s$^{-1}$. The outflow can be associated with either this AGN or the starburst. 
\\
\\
(f) The [\ion{N}{2}] 205 \mic\ line shows a deficit relative to the FIR continuum, similar to the [\ion{C}{2}] deficit found in ULIRGs. This is surprising because [\ion{N}{2}] arises in the \ion{H}{2} regions and not in PDRs. This deficit is consistent with recent results from a survey conducted by \herschel--PACS, in which deficits were found for the [\ion{N}{2}] 122 \mic\ line for several ULIRGS. The line deficit in the \ion{H}{2} regions could be a result of higher ionization parameter in ULIRGs compared to more quiescent systems. The observed line ratio of [\ion{C}{2}]/[\ion{N}{2}] constrains the percentage of [\ion{C}{2}] emission in PDRs compared to \ion{H}{2} regions to be between 85\% -- 75\%.
\\
\\
(g) The ratio of $N_\mathrm{C}/N_\mathrm{CO}$ is close to unity, which is high compared to the Galactic PDRs, but consistent with the correlation between [\ion{C}{1}]/CO line strength and FIR luminosity found using data from several nearby luminous galaxies with starbursts and AGN.
\\
\\
In summary, this is the first time we have a comprehensive picture of the state of molecular gas in Arp~220. We find that the molecular gas is highly influenced by the mechanical energy provided by the on-going merger. It is important to learn more about the geometry of this galaxy to see how the warm gas and the outflow are distributed. Observations from ALMA will be able to provide this information.
\acknowledgements

Acknowledgements:  We thank our anonymous referee for constructive comments that have strengthened this paper. We are grateful to the FTS/ICC team for helping us with the FTS data reduction and understanding the instrumental effects. We would like to thank Harshal Gupta from JPL for helping us with important line idenitfications. 
SPIRE has been developed by a consortium of institutes led by Cardiff Univ. (UK) and including Univ. Lethbridge (Canada); NAOC (China); CEA, LAM (France); IFSI, Univ. Padua (Italy); IAC (Spain); Stockholm Observatory (Sweden); Imperial College London, RAL, UCL-MSSL, UKATC, Univ. Sussex (UK); Caltech, JPL, NHSC, Univ. Colorado (USA). This development has been supported by national funding agencies: CSA (Canada); NAOC (China); CEA, CNES, CNRS (France); ASI (Italy); MCINN (Spain); SNSB (Sweden); STFC, UKSA (UK); and NASA (USA).

\newpage
\begin{figure}
\begin{center}
\includegraphics[scale=0.45]{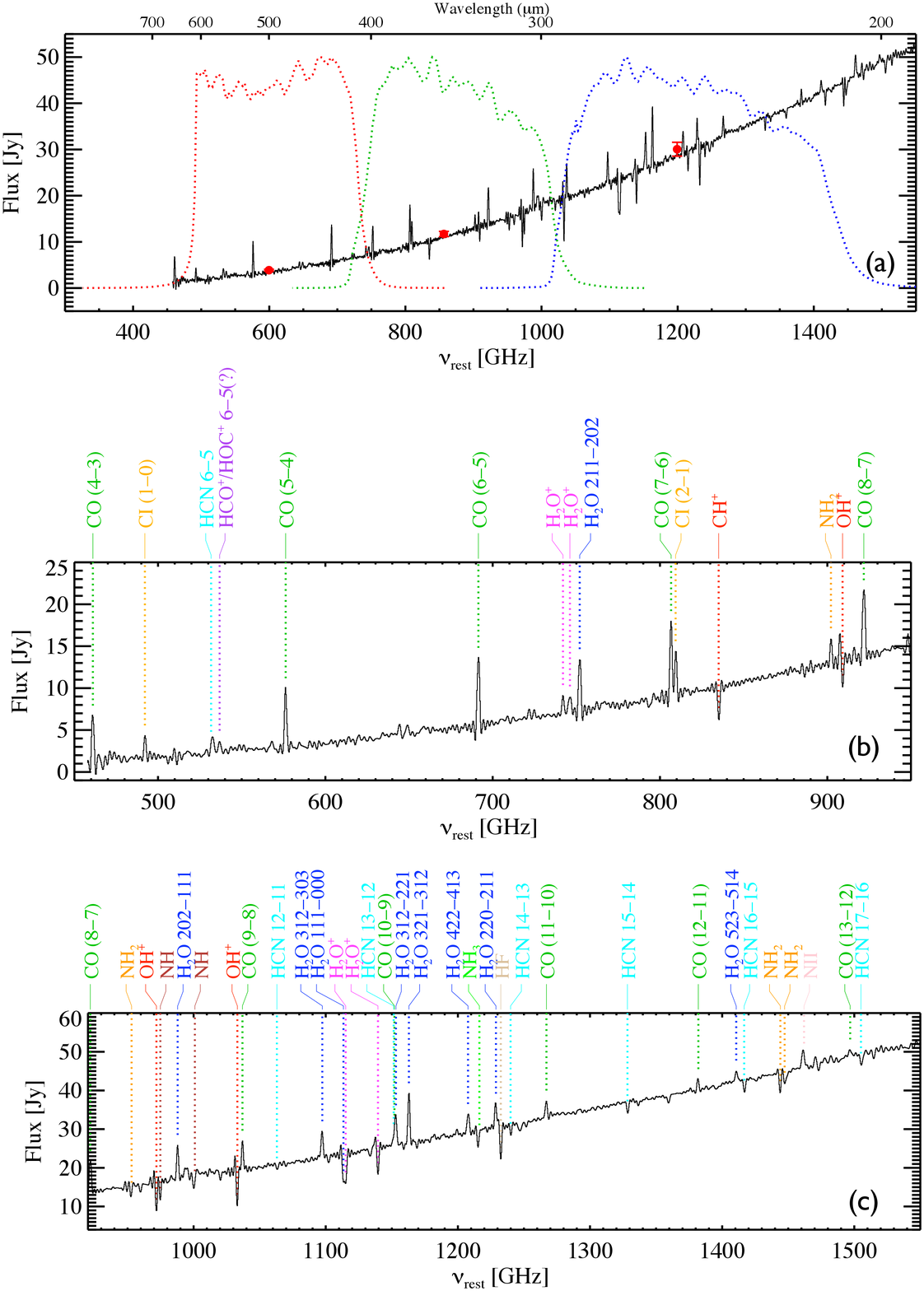}
\caption{Herschel SPIRE-FTS spectrum of Arp~220.  The spectrum shows the FIR continuum between 190 -- 670 \mic\ in (a). Red solid points in (a) are the continuum measurements from SPIRE photometer and the dotted curves show the photometer bandpasses with arbitrary normalization. Line identifications are shown for several molecular and atomic species in (b) and (c) with like colors for like species.} 
\label{spec}
\end{center}
\end{figure}

\begin{figure}
\begin{center}
\includegraphics[scale=0.65,angle=90]{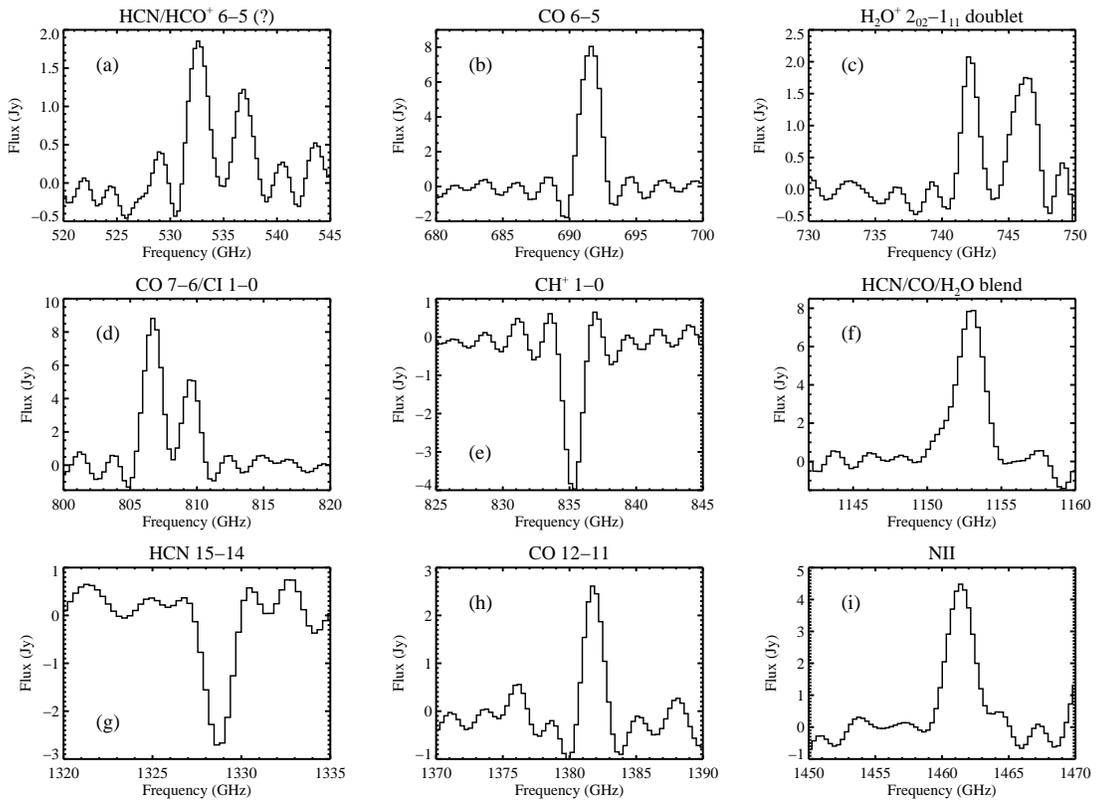}
\caption{A zoom-in view of some of the spectral lines detected in the Herschel-SPIRE spectrum of Arp~220.}
\label{specmulti}
\end{center}
\end{figure}

\begin{figure}
\begin{center}
\includegraphics[scale=0.5]{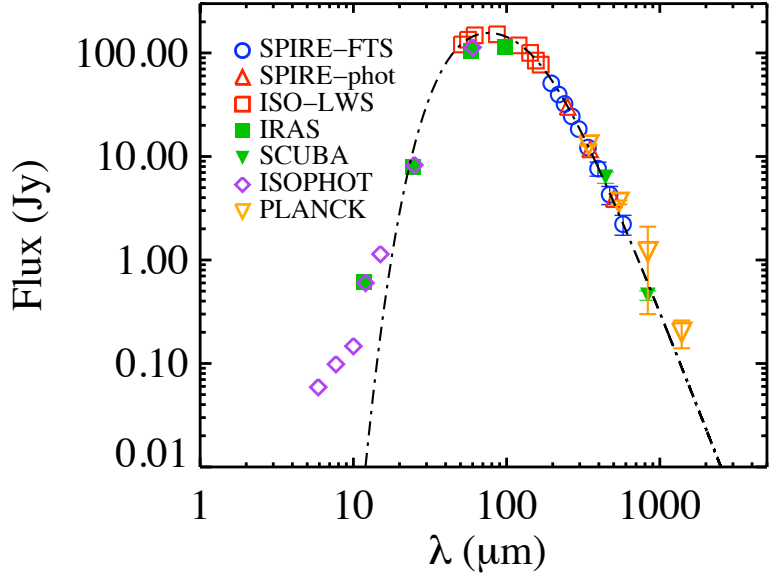}
\caption{Dust spectral energy distribution of Arp~220: The dot-dashed line is a modified blackbody model fit to the combined FTS, SPIRE-photometer and ISO-LWS data. Other data from the literature are over plotted for comparison and agree well with the fit. A single temperature component model fits the data extremely well down to 15 $\mu$m, below which it breaks down. The size of the data points is approximately equal to the size of the uncertainties (random and systematic).}
\label{dustsed}
\end{center}
\end{figure}

\begin{figure*}
\begin{center}
\includegraphics[scale=0.5]{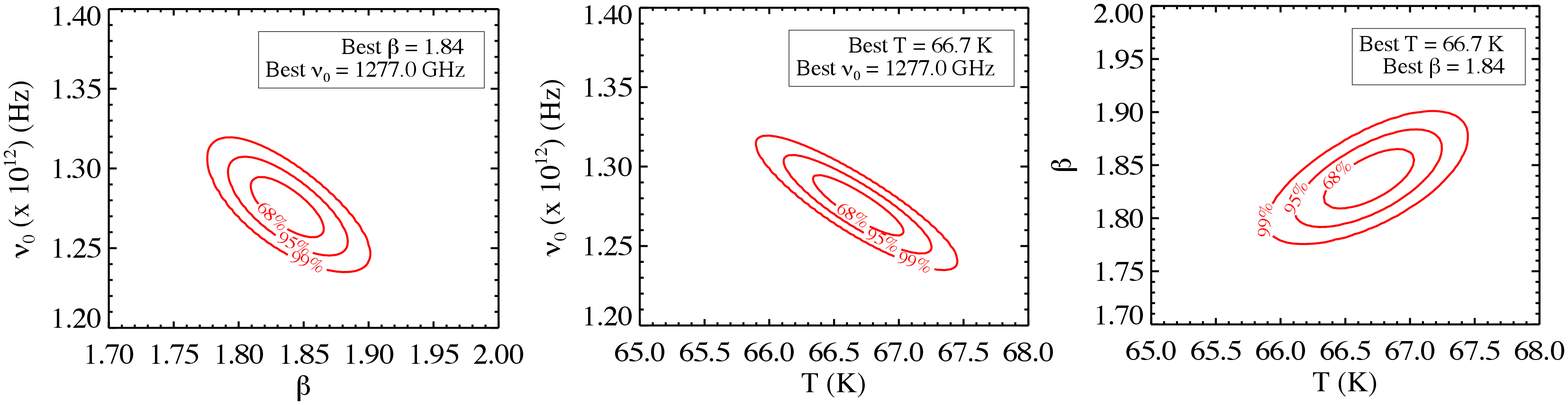}
\caption{Modeling the dust SED yields tight constraints on a single-component dust spectrum in Arp~220. The contours show 68\%, 95\% and 98\% confidence levels.}
\label{dust}
\end{center}
\end{figure*}

\begin{figure}
\begin{center}
\includegraphics[scale=0.5]{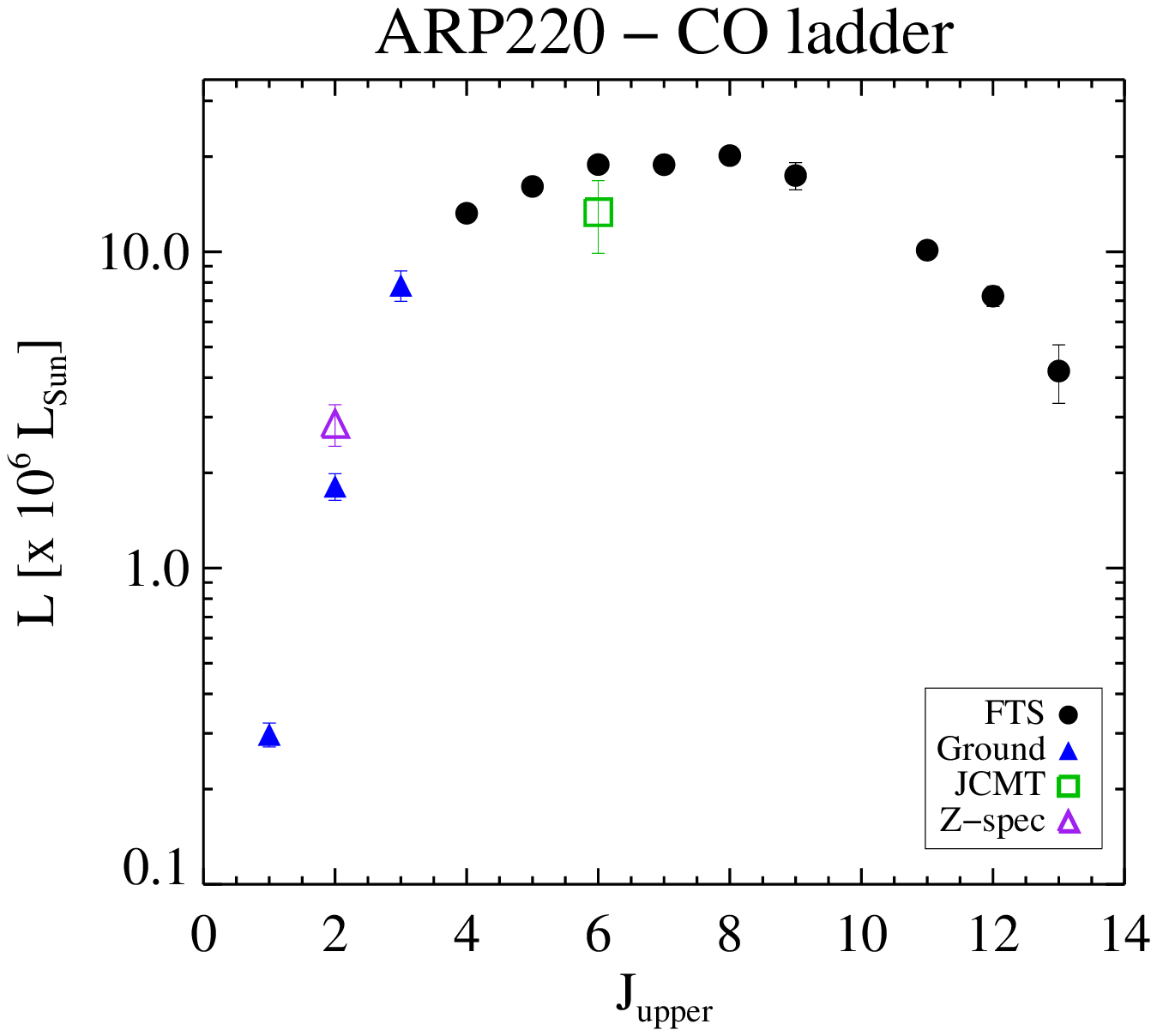}
\includegraphics[scale=0.55]{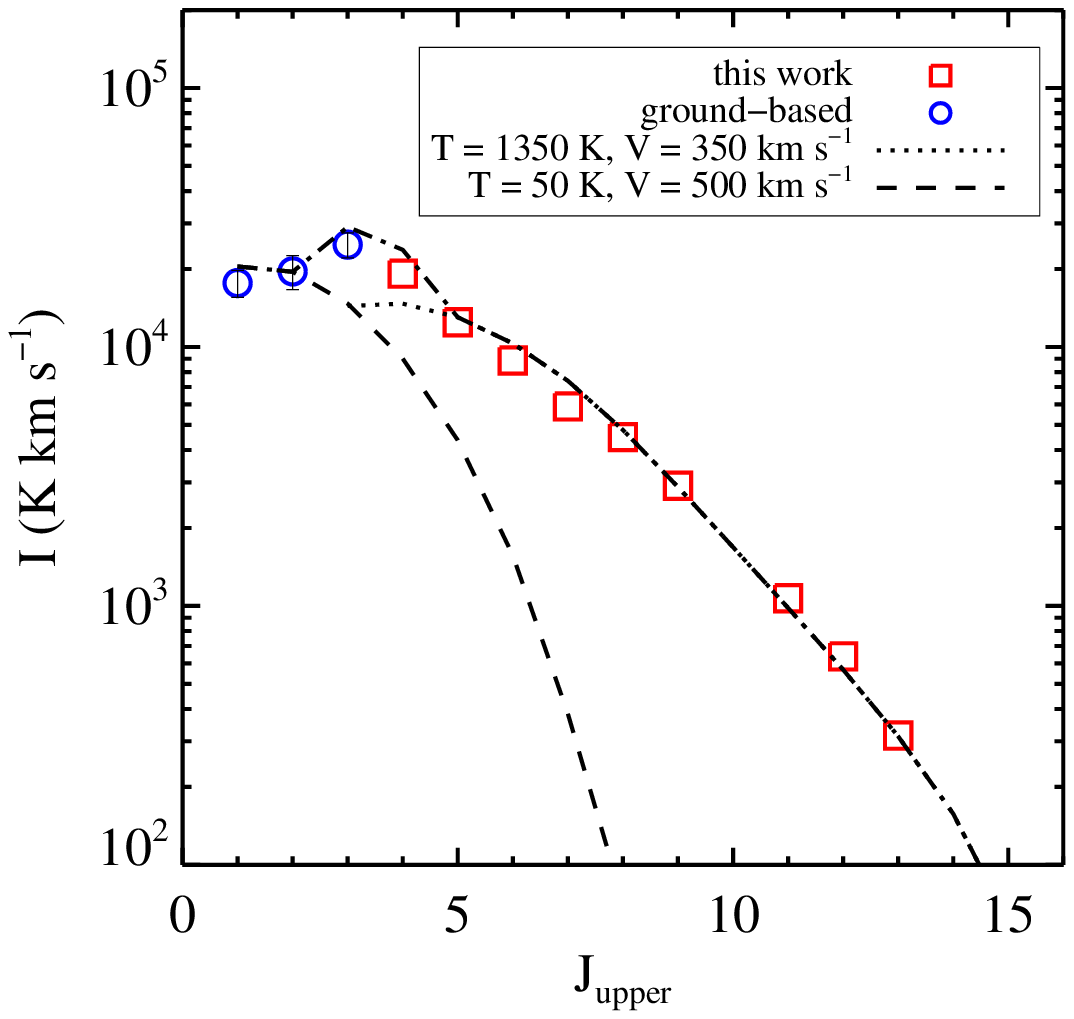}
\caption{Left: Extinction corrected luminosity distribution of the CO ladder from J = 1-0 to J = 13-12 with the exception of J = 10-9 line, which is blended with a water line. The black solid circles are FTS measurements and the blue triangles are average line fluxes from  ground-based measurements. Also shown for comparison are new measurements by Z-spec and JCMT. Right: Non-LTE model fit to the observed CO line temperatures. There are two temperature components: a warm component (dotted line) traced by the mid-J to high-J lines and a cold component (dashed line)  traced by the low-J lines.}
\label{cosed}
\end{center}
\end{figure}

\begin{figure*}
\begin{center}
\includegraphics[scale=0.5]{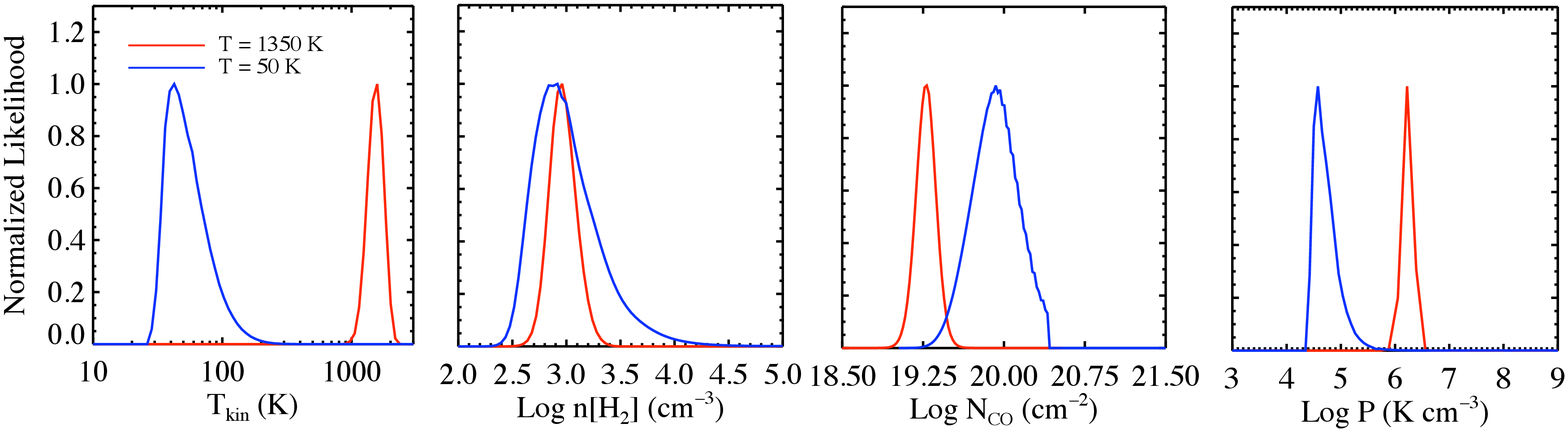}
\caption{Radiative transfer modeling of CO: Likelihood distributions for gas kinetic temperature, density, column density and pressure. The blue line represents the cold component at 50 K obtained from modeling low-J transitions and the red line represents the warm component at 1350 K obtained from modeling the mid-J to high-J transitions. }
\label{co}
\end{center}
\end{figure*}

\begin{figure*}
\begin{center}
\includegraphics[scale=0.5]{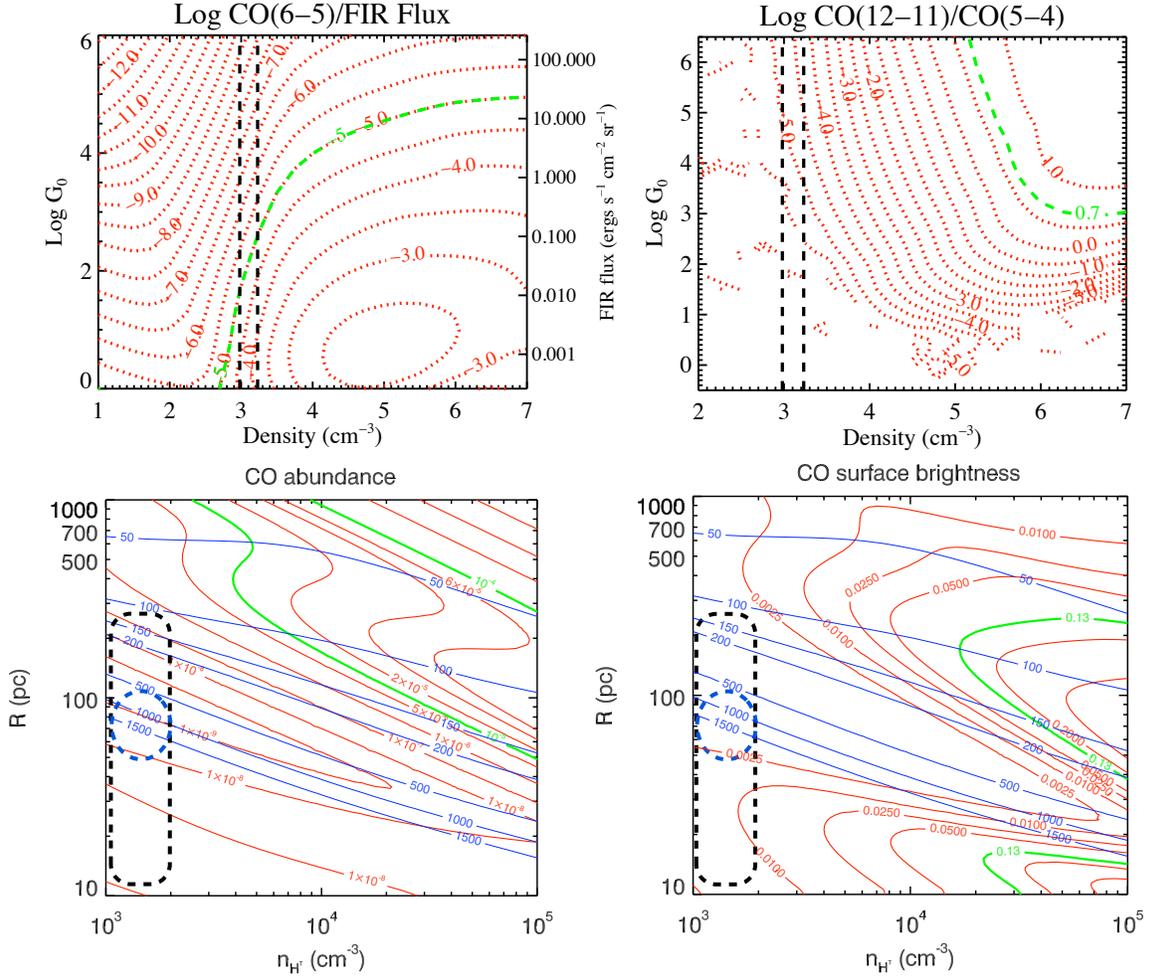}
\caption{Top: PDR models for CO \citep{wolfire10}. The red lines are model contours, green line is the observed value (except for the CO abundance for which it represents an expected value derived from Galactic observations), and the black dashed lines are acceptable density limits from non-LTE modeling of CO. Bottom: XDR models for CO abundance and CO surface Brightness in erg \cmtwo\ s$^{-1}$ sr$^{-1}$. The red lines are model contours, blue lines are temperature contours, green lines are observed values, black dashed region shows acceptable ranges for the source size and density from Table~3, and the blue dash circle shows the $3-\sigma$ limits for the temperature as listed in Table~3. }
\label{pdrxdr}
\end{center}
\end{figure*}

\begin{figure}
\begin{center}
\includegraphics[scale=0.45]{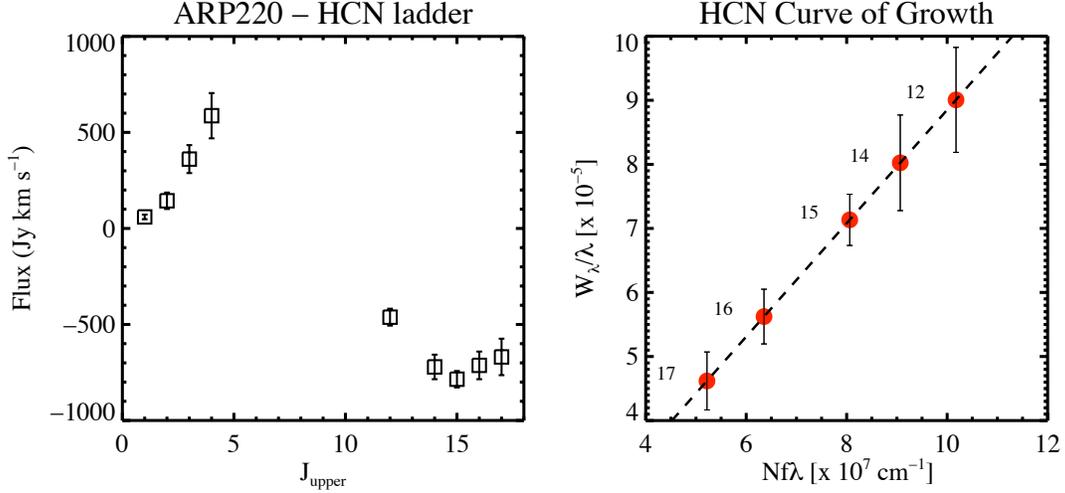}
\caption{Left -- HCN rotational ladder: The low-J lines measured from the ground are in emission, while the high-J lines measured by SPIRE-FTS are in absorption. Right -- HCN curve of growth: All the HCN absorption lines from J = 12-11 to J = 17-16 are on the linear part of the curve of growth.}
\label{hcncog}
\end{center}
\end{figure}

\begin{figure*}
\begin{center}
\includegraphics[scale=0.7]{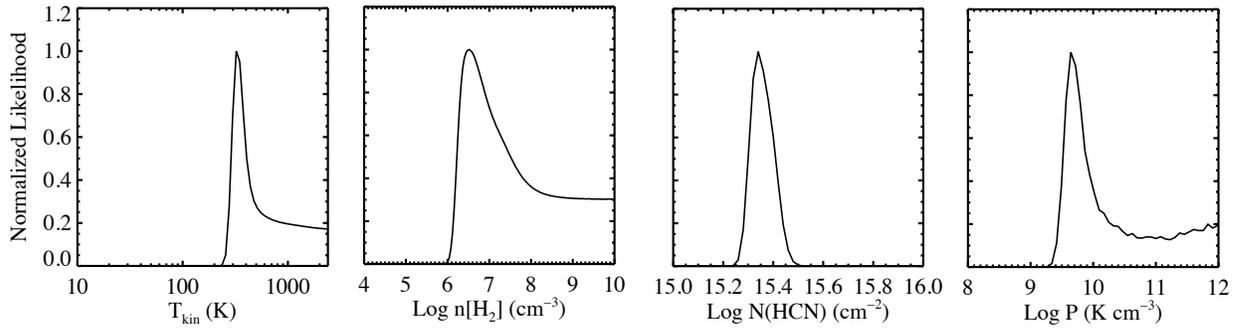}
\caption{Radiative transfer modeling of HCN: Likelihoods for gas kinetic temperature, density, column density and pressure. These results do not include the effects of IR pumping. }
\label{hcnlike}
\end{center}
\end{figure*}

\begin{figure*}
\begin{center}
\includegraphics[scale=0.55]{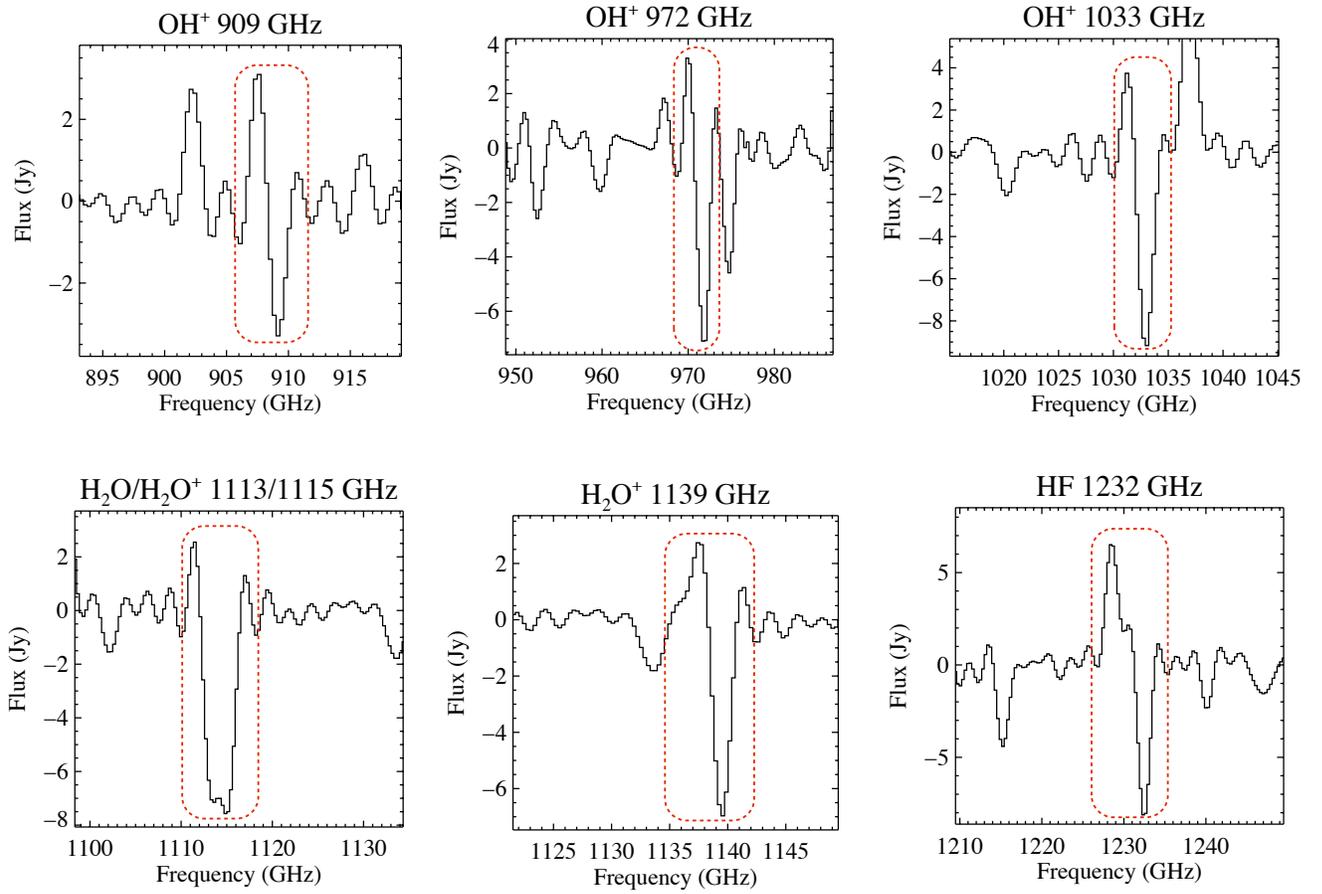}
\caption{P Cygni detected in OH+, H2O+ and HF lines, suggesting molecular outflow in Arp~220. The dashed region highlights the P Cygni profile of the molecule.}
\label{pcygni}
\end{center}
\end{figure*}

\begin{figure}
\begin{center}
\includegraphics[scale=0.45]{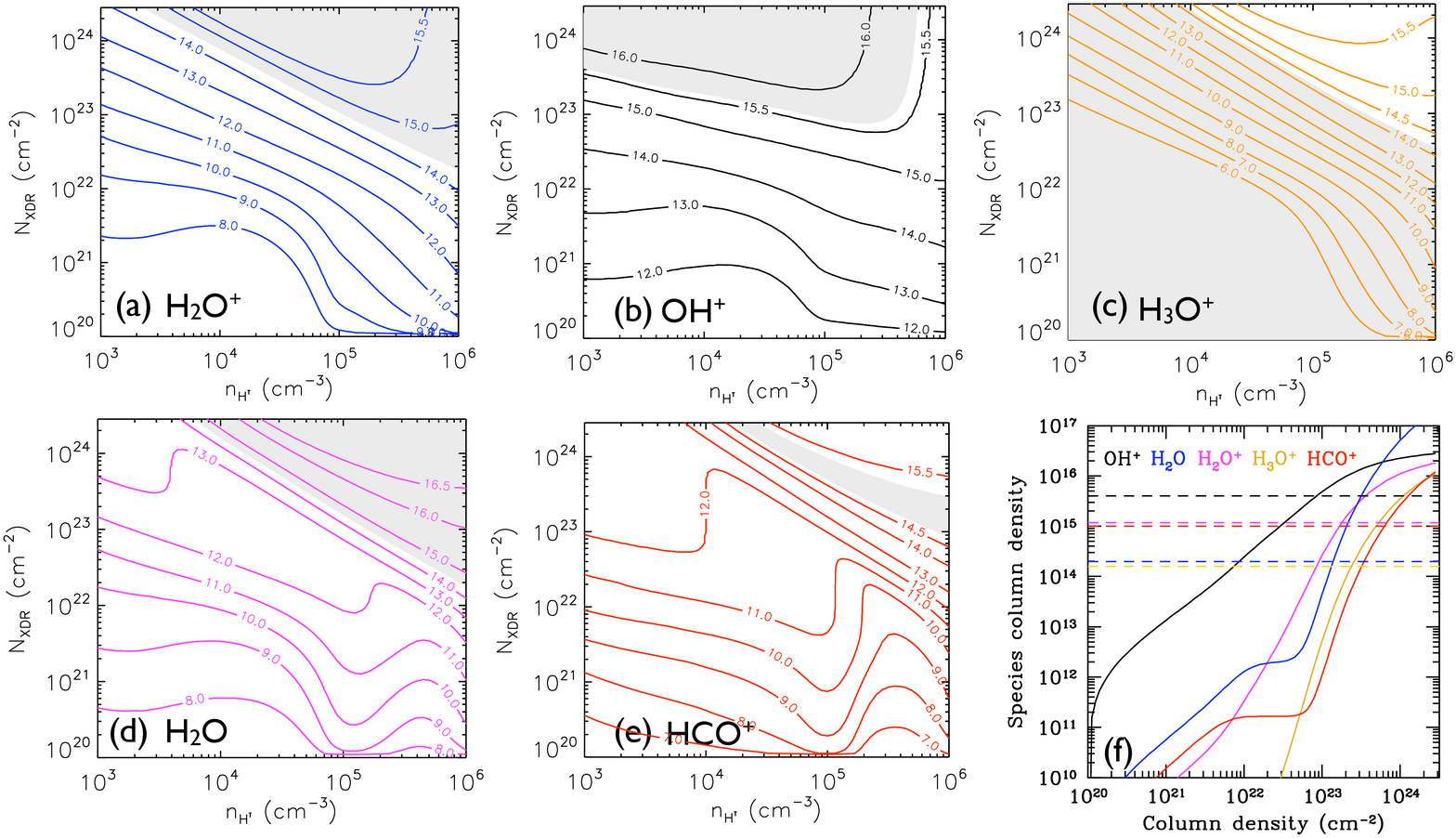}
\caption{XDR Model: (a) -- (e) Predictions for columns densities of several ion species over the $n_\mathrm{H}$ and $N_\mathrm{H}$ grid. The shaded area shows the allowed regions of column densities for OH$^{+}$, \water$^{+}$ and \water\, H$_3$O$^{+}$ and HCO$^{+}$. (f) -- Total columns densities of each species as a function of total hydrogen column. The dashed lines are observed lower limits on the column densities for all molecules except for HCO$^{+}$ and  H$_3$O$^{+}$, for which they represent an observed value and a $3-\sigma$ upper limit, respectively.}
\label{xdr}
\end{center}
\end{figure}

\newpage
\begin{deluxetable}{clllcl}
\tabletypesize{\scriptsize} \tablecaption{Line Strengths of Emission and Absorption Lines in Arp 220}
\tablehead{\multicolumn{6}{c}{Emission Lines}\\
\cline{1-6}\\
 \colhead{ID}& \colhead{Transition} & \colhead{Rest Frequency} & \colhead{Flux}  & \colhead{Source} & \colhead{Comments}\\
\colhead{} & \colhead{}& \colhead{(GHz)} & \colhead{(Jy km s$^{-1})$} & \colhead{} & \colhead{}}
\startdata
\coone\  &  J = 4 - 3 & 461.04077 & 4550 $\pm$ 330   & JPL/CDMS & -- \\
\coone\  &  J = 5 - 4 &  576.26793 & 3660 $\pm$ 170   & JPL/CDMS & --\\
\coone\ &  J = 6 - 5 &  691.47308 & 4070 $\pm$ 80       & JPL/CDMS & -- \\
\coone\  &  J = 7 - 6 &  806.65181 & 3460  $\pm$ 170   & JPL/CDMS & -- \\
\coone\ &  J = 8 - 7 &  921.79970 &  3260  $\pm$ 200   & JPL/CDMS & -- \\
\coone\  &  J = 9 - 8 & 1036.91239 &  2920 $\pm$  270  & JPL/CDMS & -- \\
\coone\ &  J = 10 - 9 &  1151.98545 &  ... &  JPL/CDMS & blended with \water\ $3_{12} - 2_{21}$\\
\coone\  &  J = 11-10 & 1267.01449 &  1280 $\pm$ 90   &  JPL/CDMS & -- \\
\coone\  &  J = 12-11 & 1381.99510 & 840  $ \pm$ 60    & JPL/CDMS & -- \\
\coone\  &  J = 13-12  &  1496.92291 & 450 $\pm$ 100     &  JPL/CDMS & -- \\
CI    &   J = 1-0  &   492.16065   &  1840 $\pm$ 70 & CDMS & -- \\ 
CI    &    J = 2-1  &   809.34197   & 2130 $\pm$  400 & CDMS & --\\
\water\ & $2_{11} - 2_{02}$ & 752.0332  & 2970 $\pm$ 110 & LOVAS & -- \\
\water\ & $2_{02} - 1_{11}$ & 987.92676 & 3440 $\pm$ 450 &  JPL & -- \\
\water\ & $3_{12} - 3_{03}$ & 1097.36479 & 3080 $\pm$ 160 & JPL & -- \\
\water\ & $3_{12} - 2_{21}$ & 1153.12682 & 3460 $\pm$ 330 & JPL & blended with CO J = 10-9 line\\
\water\ & $3_{21} - 3_{12}$  & 1162.91159 & 4930 $\pm$ 300 & JPL & -- \\ 
\water\ & $4_{22} - 4_{13}$  & 1207.63871 & 1750 $\pm$ 170 & JPL & --\\
\water\ & $2_{20} - 2_{11}$  & 1228.78877  & 1980 $\pm$ 300 & JPL & --\\
\water\ & $5_{23} - 5_{14}$  & 1410.61807  & 940 $\pm$ 110 &   JPL & -- \\
\water$^{+}$ & $2_{02}-1_{11}$, $J_{5/3-3/2}$ & 742.0332 & 840 $\pm$ 90 & BR10\tablenotemark{b} & -- \\
\water$^{+}$ & $2_{02}-1_{11}$, $J_{3/2-3/2}$ & 746.1938 & 980 $\pm$ 130 & BR10/JPL & -- \\
HCN & J = 6-5 &  531.71635 & 1700 $\pm$ 120 & CDMS & (?)\\
HCO$^{+}$/HOC$^{+}$ & J = 6-5 & 536.82796 & 850 $\pm$ 160 & CDMS & (?)\\
NII  &  $^{3}P_{1} - ^{3}P_{0}$ & 1461.1319 & 1800 $\pm$ 120  & C93\tablenotemark{c} & very broad  \\
\cline{1-6}\\
\multicolumn{6}{c}{Absorption lines}\\
\cline{1-6}\\
\colhead{ID}& \colhead{Transition} & \colhead{Rest Frequency} & \colhead{W$_{\lambda}$}  & \colhead{Source} & \colhead{Comments}\\
\colhead{} & \colhead{}& \colhead{(GHz)} & \colhead{($\times 10^{-2} \mu$m)} & \colhead{} & \colhead{}\\
\cline{1-6}\\
HCN & J = 12-11 & 1062.980  & 2.5 $\pm$ 0.2 & CDMS & --\\
HCN & J = 13-12 & 1151.449 & ... & CDMS & blended with CO (10-9) and water line\\
HCN & J = 14-13 & 1239.890 & 1.9 $\pm$ 0.2 & CDMS & --\\
HCN & J = 15-14 & 1328.302 & 1.6 $\pm$ 0.1 & CDMS & --\\
HCN & J = 16-15 & 1416.683 & 1.2 $\pm$ 0.1 & CDMS & --\\
HCN & J = 17-16 & 1505.030 & 0.9 $\pm$ 0.1 & CDMS & --\\
CH$^{+}$  & 1 - 0 & 835.07895 & 18.8 $\pm$ 0.3 & CDMS & --\\
OH$^{+}$  & $1_{01} - 0_{12}$  & 909.1588 &  11.1 $\pm$ 1.4 & CDMS & --\\ 
OH$^{+}$  & $1_{22} - 0_{11}$ & 971.8053  & 16.6 $\pm$ 2.0 & CDMS & --\\ 
OH$^{+}$  & $1_{12} - 0_{1,2}$  & 1033.118  & 16.5 $\pm$ 1.8 & CDMS & --\\
\water\ & $1_{10} - 0_{00}$ & 1113.34296 & 8.4 $\pm$ 1.4 & JPL & -- \\
\water$^{+}$ & $1_{11}-0_{00}$, $J_{3/2-1/2}$   & 1115.2040  & 9.1 $\pm$ 1.2 & G10\tablenotemark{d}  & -- \\
\water$^{+}$ & $1_{11}-0_{00}$, $J_{3=1/2-1/2}$   & 1139.5606 & 8.6 $\pm$ 0.5 & G10 & -- \\
HF & J = 1-0 & 1232.47622 & 7.0 $\pm$ 0.7 & JPL & -- \\
\enddata
\tablenotetext{a}{JPL: Jet Propulsion Lab, CDMS: Cologne Database for Molecular Spectroscopy, LOVAS: NIST recommended rest frequencies.}
\tablenotetext{b}{\citet{bruderer10}}
\tablenotetext{c}{\citet{gupta10}}
\tablenotetext{d}{\citet{brown94,colgan93}}
\tablenotetext{e}{The line fluxes can be corrected for dust extinction using the following two relations: $\tau_{d} = (\nu/\nu_{0})^{\beta}$ (where $\beta = 1.84$ and $\nu_{0} = 1270$ GHz), and the mixed dust extinction model given by $I = I_{0}(1 - e^{-\tau_{d}})/\tau_{d}$. The extinction correction factor ($I_{0}/I$) ranges from 1.076 at 450 GHz to 1.953 at 1600 GHz}

\end{deluxetable}

\begin{deluxetable}{cl}
\tabletypesize{\scriptsize} \tablecaption{Continuum fluxes}
\tablehead{\multicolumn{2}{c}{SPIRE-FTS}\\
\cline{1-2}\\
 \colhead{Wavelength}& \colhead{Flux} \\
\colhead{($\mu$m)} & \colhead{(Jy)}}
\startdata
        $586 \pm 70$ &   $2.2 \pm      0.5$     \\
       $480 \pm 46$   &  $ 4.3   \pm   0.9$      \\
       $400 \pm 32$    &  $7.6    \pm   1.2$     \\
       $345 \pm 24$     &  $12.2  \pm    1.6$       \\
       $300 \pm 18$     & $18.4    \pm   2.0$       \\
       $270 \pm 14$     & $24.4   \pm     2.2$       \\
       $240 \pm 12$      & $32.2   \pm       2.4$       \\
       $220 \pm 10$       & $39.7  \pm       2.4$      \\
       $200  \pm 8$     & $50.9   \pm       1.6$      \\
\cline{1-2}\\
\multicolumn{2}{c}{SPIRE-photometer}\\
\cline{1-2}\\
  250     &  $30.1 \pm 1.5$\\
  350       &   $11.7 \pm 0.6$\\
  500    &   $ 3.9 \pm 0.2 $\\
\enddata
\end{deluxetable}

\begin{deluxetable}{lcll}
\tabletypesize{\scriptsize}
\tablecaption{Highest Likelihood Values and Their Uncertainties from Non-LTE Modeling}
\tablehead{\multicolumn{4}{c}{CO - Warm}\\
\cline{1-4}\\
 \colhead{Quantity}& \colhead{Highest Likelihood Value} & \colhead{1$-\sigma$ range}  & \colhead{3$- \sigma$ range}
 }
\startdata
$T_{\textrm{kin}}$ (K)  & 1343 & 1247 -- 1624 & 1174 -- 1699 \\
$\mathrm{Log}_{10}\,n({\mathrm{H}_2})$ (cm$^{-3}$) & 3.2   & 3.0 -- 3.2 & 3.0 -- 3.3\\
Log$_{10} N(^{12}$CO) (cm$^{-2}$) & 19.4 & 19.4 -- 19.5 & 19.4 -- 19.6\\
Log$_{10} P$ (K cm$^{-3}$) & 6.3 & 6.2 -- 6.4 & 6.1 -- 6.4\\
$M_{\mathrm{gas}}$ ($\times 10^{8}\, M_{\sun}$) & 4.7  &  4.6 -- 6.0 & 4.3 -- 6.9  \\
$L_\mathrm{CO} (\times 10^{8}\,L_{\odot})$ & 2.0 & 1.9 -- 2.1 & 1.7 -- 2.3\\  
d$v$/d$r$ (\kms pc$^{-1}$)  & 20  & 10 -  24 & 8 -- 28\\

\cline{1-4}\\
\multicolumn{4}{c}{CO - Cold}\\
\cline{1-4}\\

$T_{\textrm{kin}}$ (K)  & 50 & 34 -- 67 & 32 -- 83 \\
$\mathrm{Log}_{10}\,n({\mathrm{H}_2})$ (cm$^{-3}$) & 2.8  & 2.6 -- 3.2 & 2.6 -- 3.40 \\
Log$_{10} N(^{12}$CO) (cm$^{-2}$) & 20.3 & 19.9 -- 20.3 & 19.8 -- 20.4\\
Log$_{10} P$ (K cm$^{-3}$) & 4.5 & 4.5 -- 4.8 & 4.4 -- 5.0\\
$M_{\mathrm{gas}}$ ($\times 10^{9}\,\mathrm{M}_{\sun}$) & 5.2  &  2.0 -- 5.4 & 1.5 -- 6.8  \\
d$v$/d$r$ (\kms pc$^{-1}$)  & 1.4 & 1.3 -  6.0 & 1.1 -- 8.5\\

\cline{1-4}\\
\multicolumn{4}{c}{HCN - high-J}\\
\cline{1-4}\\
$T_{\textrm{kin}}$ (K)  & $\gtrsim 320$ & -- & -- \\
$\mathrm{Log}_{10}\,n({\mathrm{H}_2})$ (cm$^{-3}$) & $\gtrsim 6.3$   & -- & -- \\
Log$_{10} N(^{12}$CO) (cm$^{-2}$) & 15.3 & 15.3 -- 15.4 & 15.2 -- 15.5\\
Log$_{10} P$ (K cm$^{-3}$) & $\gtrsim 9.6$ & -- & --\\

\enddata
\end{deluxetable}

\end{document}